\documentclass[aps,prd,twocolumn,groupedaddress,nofootinbib,longbibliography,preprintnumbers,floatfix]{revtex4-2}

\usepackage[normalem]{ulem}
\usepackage{bm}
\usepackage[utf8]{inputenc}
\usepackage{amsmath}
\usepackage{amssymb}
\usepackage{amsfonts}
\usepackage{upgreek}
\usepackage{subeqnarray}
\usepackage{graphicx}
\usepackage[dvipsnames]{xcolor}
\usepackage{yfonts}
\usepackage{siunitx}
\usepackage{enumitem}

\usepackage{epsfig}
\usepackage[colorlinks]{hyperref}
\usepackage{xspace}

\usepackage{tikz}
\usetikzlibrary{tikzmark}
\usetikzlibrary{positioning}

\definecolor{lime}{HTML}{A6CE39}
\DeclareRobustCommand{\orcidicon}{\hspace{-1mm}
	\begin{tikzpicture}
	\draw[lime, fill=lime] (0,0) 
	circle [radius=0.16] 
	node[white] {{\fontfamily{qag}\selectfont \tiny \,ID}};
	\draw[white, fill=white] (-0.0525,0.095) 
	circle [radius=0.007];
	\end{tikzpicture}
	\hspace{-3mm}
}

\foreach \x in {A, ..., Z}{\expandafter\xdef\csname orcid\x\endcsname{\noexpand\href{https://orcid.org/\csname orcidauthor\x\endcsname}
			{\noexpand\orcidicon}}
}

\hypersetup{linkcolor=BrickRed,citecolor=Green,
filecolor=Mulberry,
urlcolor=NavyBlue,
menucolor=BrickRed,
runcolor=Mulberry
}

\newcommand{\be}{\begin{equation}}
\newcommand{\ee}{\end{equation}}
\newcommand{\bea}{\begin{eqnarray}}
\newcommand{\eea}{\end{eqnarray}}
\newcommand{\beas}{\begin{subeqnarray}}
\newcommand{\eeas}{\end{subeqnarray}}

\newcommand{\primat}{\texttt{PRIMAT}\xspace}
\newcommand{\parthenope}{\texttt{PArthENoPE}\xspace}
\newcommand{\nevo}{\texttt{NEVO}\xspace}

\newcommand{\allCMB}{\text{all-CMB}}

\makeatletter
\newcommand{\abs}{\@ifstar\abssmall\absbig}
\newcommand{\absbig}[1]{\left \lvert #1 \right \rvert}
\newcommand{\abssmall}[1]{\lvert #1 \rvert}
\makeatother

\newcommand{\vrho}{\varrho}
\newcommand{\bvrho}{\bar{\varrho}}

\newcommand{\Neff}{N_{\mathrm{eff}}}
\newcommand{\Neffxi}{N_{\mathrm{eff} \, (\xi_\alpha)}}
\newcommand{\Yp}{Y_\mathrm{p}}
\renewcommand{\DH}{\mathrm{D}/\mathrm{H}}

\newcommand{\Hself}{\mathcal{J}}

\renewcommand{\H}{\mathcal{H}}
\newcommand{\ATAO}{{ATAO}}
\newcommand{\ATAOH}{{ATAO}-$\mathcal{V}\,$}

\newcommand{\ATAOJH}{{ATAO}-$(\Hself\pm\mathcal{V})\,$}

\newcommand{\obs}{\mathrm{obs}}
\newcommand{\code}{\mathrm{code}}

\newcommand{\dd}{\mathrm{d}}

\newcommand{\com}[1]{}
\newcommand{\ii}{\mathrm{i}}
\newcommand{\Tcm}{T_{\mathrm{cm}}}
\newcommand{\xit}{\tilde{\xi}}
\newcommand{\av}{\mathrm{av}}
\newcommand{\eqb}{\mathrm{mix}}
\newcommand{\trans}{\mathrm{trans}}
\newcommand{\Deltat}{\widetilde{\Delta}}

\renewcommand{\L}{\mathcal{L}}
\renewcommand{\vec}{\mathbf}

\begin{document}
\preprint{N3AS-24-015}

\title{Constraints on primordial lepton asymmetries with full neutrino transport}

\author{Julien Froustey\orcidA{}}
\email[]{jfroustey@berkeley.edu}
\affiliation{Department of Physics, University of California Berkeley, Berkeley, California 94720, USA}
\affiliation{Department of Physics, North Carolina State University, Raleigh, North Carolina 27695, USA}

\author{Cyril Pitrou\orcidB{}}
\email[]{pitrou@iap.fr}
\affiliation{Institut d'Astrophysique de Paris, CNRS UMR 7095,
Sorbonne Université, 98 bis Bd Arago, 75014 Paris, France}

\begin{abstract}
Primordial neutrino-antineutrino asymmetries can be constrained through big-bang nucleosynthesis (BBN) relic abundances and cosmic microwave background (CMB) anisotropies, both observables being sensitive to neutrino properties. The latter constraint, which is due to gravitational effects from all neutrino flavors, is very minute since it is at least quadratic in the asymmetries. On the contrary, the constraints from primordial abundances presently dominate, although these abundances are almost only sensitive to the electron flavor asymmetry. It is generally assumed that neutrino asymmetries are sufficiently averaged by flavor oscillations prior to BBN, which allows one to constrain a common primordial neutrino asymmetry at the epoch of BBN. This simplified approach suffers two caveats that we deal with in this article, combining a neutrino evolution code and BBN calculation throughout the MeV era. First, flavor “equilibration” is not true in general, therefore an accurate dynamical evolution of asymmetries is needed to connect experimental observables to the \emph{primordial} asymmetries. Second, the approximate averaging of asymmetries through flavor oscillations is associated to a reheating of the primordial plasma. It is therefore crucial to correctly describe the interplay between flavor equilibration and neutrino decoupling, as an energy redistribution prior to decoupling does not significantly alter the final effective number of neutrino species' value. Overall, we find that the space of allowed initial asymmetries is generically unbound when using currently available primordial abundances and CMB measurements. We forecast constraints using future CMB experiment capabilities, which should reverse this experimental misfortune.
\end{abstract}

\maketitle

\section{Introduction}

Neutrinos play a major role in various stages of the history of the Universe, for instance through their effect on the primordial abundances of elements produced during Big Bang Nucleosynthesis (BBN), on the power spectrum of Cosmic Microwave Background (CMB) anisotropies, or on cosmological structure formation~\cite{NeutrinoCosmology}. Consequently, the Universe can be seen as a vast “laboratory” for exploring neutrino physics. In particular, the phenomenon of neutrino flavor oscillations is a direct evidence of the need for beyond-the-Standard-Model physics. The effect of flavor oscillations on neutrino evolution is particularly relevant in the early Universe when one considers non-zero asymmetries, i.e., differences between the distributions of neutrinos and antineutrinos.

The lepton asymmetry of the Universe is very loosely constrained compared to the baryon asymmetry, $(n_b - \bar{n}_b)/n_\gamma = (6.10 \pm 0.4) \times 10^{-10}$~\cite{Planck:2018vyg}, where $n_b$ (resp. $\bar{n}_b$) is the total number density of baryons (resp. antibaryons), and $n_\gamma$ the photon number density. This baryon asymmetry is believed to originate from a dynamical process (“baryogenesis”). Because of sphaleron processes in the very early Universe, one would expect baryon and lepton asymmetries to be of the same order of magnitude (see e.g.,~\cite{Harvey:1990qw,Dreiner:1992vm}). However, many models have been constructed with a large relic lepton asymmetry compared to the baryon asymmetry (e.g.,~\cite{March-Russell:1999hpw,McDonald:1999in,Pilaftsis:2003gt,Asaka:2005pn,Canetti:2012kh,Ghiglieri:2020ulj,Kawasaki:2022hvx,Borah:2022uos}). In addition, non-zero neutrino asymmetries are promising ways to tackle some issues in cosmology: for instance, they allow one to relax constraints on the thermalization of sterile neutrinos~\cite{Foot:1995bm,Abazajian:2004aj,Chu:2006ua,Hannestad:2012ky,Mirizzi:2012we,Saviano:2013ktj}). Extended $\Lambda$CDM models including a neutrino chemical potential have also been shown to be able to alleviate cosmological tensions~\cite{Barenboim:2016lxv,Yeung:2020zde,Seto:2021tad,Kumar:2022vee,Yeung:2024krv}.

An asymmetry between electron neutrinos and antineutrinos changes the equilibrium neutron-to-proton ratio prior to BBN, which results in modifications of the primordial abundances. In addition, the energy density of (anti)neutrinos is modified with respect to the standard, symmetric case. As a consequence, spectroscopic measurements of the primordial abundances and analysis of CMB anisotropies, which allow one to determine the expansion history of the Universe, can be used to constrain primordial asymmetries. Several studies have therefore evaluated the impact of including non-zero neutrino chemical potentials at the BBN epoch—e.g.,~\cite{Simha:2008mt,Shimon:2010ug,Oldengott:2017tzj,Pitrou_2018PhysRept,Matsumoto:2022tlr,Burns:2022hkq,Escudero:2022okz}—but ignoring earlier asymmetry evolution.

This approach is not fully satisfying. Indeed, because of flavor oscillations that typically become active for temperatures $T \leq \SI{10}{\mega \electronvolt}$, the connection between BBN-epoch and primordial (i.e., at temperatures $\geq \SI{20}{\mega \electronvolt}$) asymmetries is far from obvious~\cite{Savage:1990by}. In the last decades, major progress has been made toward a precise and accurate description of neutrino evolution in the MeV age, including the effect of flavor oscillations and QED corrections to the plasma thermodynamics~\cite{Dolgov_NuPhB1997,Mangano2002,Mangano2005,Grohs2015,Relic2016_revisited,Akita2020,Froustey2020,Bennett2021}, which notably led to the standard prediction for the effective number of neutrino species, $\Neff = 3.044$.\footnote{Recently, QED corrections to the interaction vertices, not considered in the calculations of \cite{Froustey2020,Bennett2021}, were estimated to reduce this number to $3.043$~\cite{Cielo:2023bqp}. However, subsequent studies dispute this conclusion~\cite{Jackson:2023zkl,Drewes:2024wbw}.} These studies did not include non-zero asymmetries, which, combined with flavor mixing, give rise to collective phenomena called “synchronous” oscillations~\cite{Samuel:1995ri,Bell98,Pastor:2001iu,Abazajian2002,Wong2002}. Subsequently, various studies have focused on the impact of flavor oscillations on the evolution of asymmetries~\cite{Dolgov_NuPhB2002,Pastor:2008ti,Mangano:2010ei,Mangano:2011ip,Castorina2012,Barenboim:2016shh,Johns:2016enc}, also assessing the role of the $CP$-phase~\cite{Gava:2010kz,Barenboim:2016jxn}. In~\cite{Froustey2021}, we developed the first three-flavor multi-momentum neutrino evolution code with asymmetries that uses the full collision term, instead of the damping approximation traditionally used for numerical efficiency. We showed that using the complete collision term is crucial since oscillations are “overdamped” when using a damping approximation, and asymmetries are not generally equilibrated to the same value, hence affecting the connection between primordial parameters and cosmological observables.

We are now in a position to revisit the problem of constraining primordial asymmetries using BBN and CMB data, not restricting to equal asymmetries or values set directly at the BBN epoch, and using the actual Standard Model collision kernel (in the low-energy Fermi approximation, valid for temperatures very small compared to $m_{W,Z}$). To this end, we will use the measurements of the helium-4 and deuterium primordial abundances, along with existing CMB anisotropies and baryon acoustic oscillations (BAO) data. In addition, following the strategy of~\cite{Escudero:2022okz}, we will forecast how these constraints will be tightened with future CMB experiments. Note that we will call “primordial” asymmetries the neutrino reduced chemical potentials at a temperature $\sim \SI{25}{\mega \electronvolt}$, regardless of how they were produced.\footnote{In particular, a recent work~\cite{Domcke:2022uue} has shown that, because of a chiral plasma instability, any reduced chemical potential $\abs{\mu_\nu}/T$ must be lower than $10^{-2}$ at a temperature of $\SI{e6}{\giga \electronvolt}$ — this is a tight constraint that can yet be avoided if asymmetries are generated below $\SI{e5}{\giga \electronvolt}$.}

This paper is organized as follows. In Sec.~\ref{sec:methods}, we introduce the various elements of our numerical and statistical analysis: quantum kinetic equations, BBN calculation and likelihood construction (additional details are gathered in Appendix~\ref{App:statistics}). In Sec.~\ref{Sec:equilibration}, we discuss the physics of flavor asymmetry equilibration and the associated thermodynamic effects in an approximate setup, which provides additional insight on our numerical methods. Our main results on the constraints of primordial asymmetries with current experimental data, in different sub-regions of the parameter space, are presented in Sec.~\ref{section:constraints}. We forecast the future capabilities of CMB experiments with regard to constraining lepton asymmetries in Sec.~\ref{sec:forecast}. Our conclusions are presented in Sec.~\ref{sec:conclusion}. Useful thermodynamic identities are gathered in Appendix~\ref{App:thermo}. In Appendix~\ref{App:Ttrans}, we discuss the convergence of our resolution scheme for the evolution of (anti)neutrinos. In Appendix~\ref{App:deuterium}, we verify that our conclusions are robust, whether the deuterium abundance (which is predicted differently by different BBN codes as they use different deuterium destruction rates) is included or not.

Throughout this paper, we work in natural units where $\hbar = c = k_B = 1$.

\section{Methods}
\label{sec:methods}

As is customary for the study of neutrino decoupling in the early Universe, we use the dimensionless variables
\begin{equation}
    x \equiv \frac{m_e}{\Tcm} \ , \quad y \equiv \frac{p}{\Tcm} \ , \quad z \equiv \frac{T_\gamma}{\Tcm} \, ,
\end{equation}
with $m_e \simeq \SI{0.511}{\mega \electronvolt}$ the electron mass, $p = \abs{\vec{p}}$ the momentum amplitude (which corresponds to the energy for ultrarelativistic particles like neutrinos), and $T_\gamma$ the photon temperature. The comoving temperature $\Tcm \propto a^{-1}$, with $a$ the scale factor, is the temperature of instantaneously decoupled neutrinos~\cite{Grohs2015}. For vanishing initial asymmetries, $\Tcm$ differs from the actual temperature of neutrinos after decoupling by $\sim 0.1 \, \%$~\cite{Froustey2020}.

The ensemble of (anti)neutrinos is described by the one-body reduced density matrices $\vrho(t,p)$, $\bvrho(t,p)$~\cite{SiglRaffelt,Volpe_2013,Froustey2020}. Given the homogeneity and isotropy of the early Universe, they only depend on the magnitude of neutrino momenta $p = \abs{\vec{p}}$. They are $3 \times 3$ matrices in flavor space; the on-diagonal components generalize the classical distribution functions, while off-diagonal elements account for flavor coherence. Using comoving variables, we write them as functions of $x$ and $y$.
 
For a given flavor $\alpha$, we characterize the asymmetry by the quantity
\begin{equation}
    \label{eq:def_asymmetry}
    \eta_\alpha \equiv \frac{n_\alpha - \bar{n}_\alpha}{\Tcm^3} = \int{\frac{y^2 \dd y}{2 \pi^2}\left[\vrho_{\alpha \alpha}(x,y) - \bvrho_{\alpha \alpha}(x,y)\right]} \, .
\end{equation}

\subsection{Initial distributions} 

For $T_{\mathrm{cm,in}} = m_e/x_\mathrm{in} \sim \SI{25}{\mega \electronvolt}$, which will be the initial temperature we consider, the high collision rates maintain neutrinos in flavor eigenstates and at equilibrium, such that $\vrho(x_\mathrm{in},y) = \mathrm{diag}(f_{\nu_e}^\mathrm{(eq)},f_{\nu_\mu}^\mathrm{(eq)},f_{\nu_\tau}^\mathrm{(eq)})$ in the flavor basis, where the equilibrium Fermi-Dirac distributions read:
\begin{equation}
\label{eq:initial_distrib}
    f_{\nu_\alpha}^\mathrm{(eq)} \equiv \frac{1}{e^{(p-\mu_\alpha)/T_\alpha}+1} = \frac{1}{e^{y/z_\alpha - \xi_\alpha}+1} \, ,
\end{equation}
with $T_\alpha$ and $\mu_\alpha$ the initial temperature and chemical potential, with the associated dimensionless quantities:
\begin{equation}
z_\alpha \equiv \frac{T_\alpha}{\Tcm}\, , \quad \xi_\alpha \equiv \frac{\mu_\alpha}{T_\alpha} \, . 
\end{equation}
At the temperature $T_{\mathrm{cm,in}}$, neutrinos are still in thermal contact with the electromagnetic plasma and $e^- e^+$ annihilations have not yet started,\footnote{We take nonetheless into account the very small deviation $z_\mathrm{in} -1 = \mathcal{O}(10^{-5})$ due to the non-fully-relativistic nature of $e^\pm$ at $T_\mathrm{cm,in}$.} such that $z_\alpha = z_\mathrm{in} \simeq 1$. Antineutrino initial distributions are identical with opposite chemical potentials, $\bar{\xi}_\alpha = - \xi_\alpha$.

Given the distributions~\eqref{eq:initial_distrib}, the initial asymmetries can be expressed as:
\begin{equation}
\label{eq:initial_asymmetry}
    \eta_{\alpha,\mathrm{in}} = z_\alpha^3 \left(\frac{\xi_\alpha}{6} + \frac{\xi_\alpha^3}{6 \pi^2}\right) = \frac16 z_\alpha^3 \xit_\alpha \, ,
\end{equation}
where we introduce
\begin{equation}
    \label{eq:def_xit}
\xit_\alpha \equiv \xi_\alpha + \frac{\xi_\alpha^3}{\pi^2} \, ,
\end{equation}
a convenient notation for handling the non-linear dependence of the asymmetry in terms of the reduced chemical potential. Throughout this paper, $\xi_\alpha$ or $\xit_\alpha$ denote the \emph{primordial} quantities, which enter the equilibrium distribution~\eqref{eq:initial_distrib} at $T_\mathrm{cm,in} \sim \SI{25}{\mega \electronvolt}$. At later times, since neutrino decoupling is an out-of-equilibrium process, neutrino distributions deviate from pure Fermi-Dirac functions and we cannot rigorously define a chemical potential or a temperature. 

A quantity that is still defined throughout the evolution via Eq.~\eqref{eq:def_asymmetry} is $\eta_\alpha$, which coincides initially with $\xit_\alpha/6$ (up to the $\mathcal{O}(10^{-5})$ difference between $z_{\alpha}$ and $1$) according to Eq.~\eqref{eq:initial_asymmetry}. Our definition differs slightly from the one in previous literature (e.g., \cite{Pastor:2008ti,Mangano:2010ei,Mangano:2011ip,Castorina2012,Grohs:2016cuu}), where the more physical parametrization $(n_\alpha -\bar n_\alpha)/n_\gamma \propto (z_\alpha/z)^3 \tilde \xi_\alpha$ is used. However, our choice benefits from the fact that the processes taking place in the neutrino medium conserve the averaged asymmetry
\begin{equation}\label{Eq:defetabar}
\hat{\eta} \left(\{z_\alpha,\xi_\alpha \}\right) \equiv \frac{1}{3} \sum_{\alpha=e,\mu,\tau} \eta_\alpha\,.
\end{equation}

\paragraph*{Parametrization —} We introduce the following parametrization of initial degeneracy parameters:
\begin{equation}
\label{eq:parametrization}
    \left\{ \begin{aligned}
        \xit_\av &\equiv \frac{\xit_e + \xit_\mu + \xit_\tau}{3} \\
        \delta \xit_e &\equiv \xit_e - \xit_\av \\
        \widetilde{\Delta} &\equiv \frac{\xit_\mu - \xit_\tau}{2}
    \end{aligned} \right. \, \iff \, 
    \left\{ \begin{aligned}
        \xit_e &= \xit_\av + \delta \xit_e \\
        \xit_\mu &= \xit_\av - \frac{\delta \xit_e}{2} + \widetilde{\Delta} \\
        \xit_\tau &= \xit_\av - \frac{\delta \xit_e}{2} - \widetilde{\Delta} \\
    \end{aligned} \right. 
\end{equation}
In Sec.~\ref{section:constraints}, we explore different sub-spaces of the three-dimensional set $\{\xit_e,\xit_\mu,\xit_\tau\}$, through specific choices of $\xit_\av$, $\delta \xit_e$ and $\Deltat$. For instance, equal initial asymmetries among all three flavors are described by $\delta \xit_e = \Deltat = 0$.

\subsection{Neutrino transport}
\label{subsec:Nu_transport}

\subsubsection{Quantum Kinetic Equations}

We start our calculation at $T_\mathrm{cm,in} = 25 \, \mathrm{MeV}$, where we specify the set of initial asymmetries $\{\xit_e,\xit_\mu,\xit_\tau\}$ and the value of the baryon density $\omega_b$. The initial temperature ratio $z_\mathrm{in} = z_{\alpha}$ is set by solving the energy conservation equation $\dot{\rho} + 3 H (\rho + P) = 0$ with all species coupled, from $(x=0,z=1)$ to $(x_\mathrm{in},z_\mathrm{in})$. The density matrix for each momentum bin is initialized with the distributions~\eqref{eq:initial_distrib}. We then run our neutrino evolution code \nevo~\cite{Froustey2020,Froustey2021,FrousteyPhD} down to a temperature of a few $\mathrm{keV}$. This provides the final value of $\Neff$ along with the neutrino distributions throughout the evolution. \nevo solves the Quantum Kinetic Equations (QKEs) \cite{SiglRaffelt,Volpe_2013,BlaschkeCirigliano,Froustey2020,Froustey2021}:\footnote{Some studies suggest that, in some regimes, a treatment beyond the reduced single-particle level could be necessary, see e.g.,~\cite{Ho:2005vj,Patwardhan:2022mxg}.}
\begin{subequations}
\begin{align}
    \frac{\partial \vrho(x,y)}{\partial x} &= - \ii \left[\H_0 + \H_\mathrm{lep} + \Hself, \vrho \right] + \mathcal{K} \, , \label{eq:QKE_rho} \\
    \frac{\partial \bvrho(x,y)}{\partial x} &= + \ii \left[\H_0 + \H_\mathrm{lep} - \Hself, \bvrho \right] + \overline{\mathcal{K}} \, , \label{eq:QKE_rhobar}
\end{align}
\end{subequations}
where the vacuum term $\H_0$ is
\begin{equation}
    \H_0 = \frac{1}{xH} \frac{1}{\Tcm} \times U \frac{\mathbb{M}^2}{2y} U^\dagger \, ,
\end{equation}
with $H$ the expansion rate, $\mathbb{M}^2 = \mathrm{diag}(0,\Delta m^2_{21},\Delta m^2_{31})$ the matrix of mass-squared differences, and $U$ the Pontecorvo-Maki-Nakagawa-Sakata matrix; the matter mean-field Hamiltonian $\H_\mathrm{lep}$ reads
\begin{equation}
\label{eq:H_matter}
    \H_\mathrm{lep} = - \frac{1}{xH} \Tcm^5 \times 2 \sqrt{2} G_F y \frac{\mathbb{E}_\mathrm{lep} + \mathbb{P}_\mathrm{lep}}{m_W^2} \, ,
\end{equation}
with $\mathbb{E}_\mathrm{lep}$ (resp.~$\mathbb{P}_\mathrm{lep}$) the diagonal matrix of charged leptons' comoving energy densities (resp.~pressures); and the self-interaction mean-field is proportional to the neutrino asymmetry matrix:
\begin{equation}
    \Hself = \frac{1}{xH} \Tcm^3 \times \sqrt{2} G_F \int{\frac{\mathrm{y}^2 \dd \mathrm{y}}{2 \pi^2}(\vrho - \bvrho)} \, .
\end{equation}
Note that we neglect a subdominant contribution similar to \eqref{eq:H_matter} involving the (anti)neutrino energy densities. Finally, the collision terms $\mathcal{K}$, $\overline{\mathcal{K}}$ account for scattering with charged leptons, scattering between (anti)neutrinos, and pair creation/annihilation reactions. The full expression of these two-dimensional collision integrals can be found in, e.g., \cite{BlaschkeCirigliano,Froustey2020,FrousteyPhD}. We compute them using their complete flavor structure, without any “damping” approximation. This is the most time-consuming part of the neutrino transport code: given the parameter sweep we perform in this study, we limit the momentum grid size to $N_y = 30$ points, linearly spaced from $y_\mathrm{min} = 0.02$ to $y_\mathrm{max} = 23$. This limited energy resolution leads to a negligible underestimation of $\Neff$. For instance, for zero asymmetries, the absolute difference between $\Neff\rvert_{N_y=30}$ and $\Neff\rvert_{N_y=80}$ is $\mathcal{O}(10^{-4})$, which is much smaller than the precision on $\Neff$ of current or future CMB experiments. The evolution of $z$ is obtained via the energy conservation equation, which includes QED corrections to the plasma thermodynamics up to order $\mathcal{O}(e^3)$~\cite{Bennett2020}.

We restrict to the normal ordering of neutrino masses, and take a zero Dirac $CP$-phase $\delta$ in the mixing matrix, as it was shown in~\cite{Froustey2021} that, in this situation, having $\delta \neq 0$ affects only marginally $\Neff$ or the spectra of $\nu_e, \bar{\nu}_e$, which are the parameters affecting CMB and BBN observables. All values for the other physical constants and mixing parameters are taken from~\cite{PDG-2022}.

Given the increase of frequency of synchronous oscillations when the temperature decreases, a solution to make the calculation numerically tractable is to use the large separation of scales between the oscillation frequencies, the collision rate, and the change rate of the Hamiltonian. It allows one to accurately describe neutrino evolution by averaging oscillations and considering that, at each time step, neutrino density matrices are diagonal in the basis where the Hamiltonian is diagonal. Adiabaticity being satisfied (i.e., the effective mixing angles varying slowly compared to the oscillation frequencies), it is sufficient to keep track of the three diagonal components of $(\vrho,\bvrho)$ in this slowly time-varying basis. One can then obtain the components of $\vrho$ in the flavor basis by using the instantaneous mixing matrix. The method based on this approximation, called “Adiabatic Transfer of Averaged Oscillations” (\ATAO), was validated in~\cite{Froustey2020} for standard neutrino decoupling and extended to include lepton asymmetries in~\cite{Froustey2021}.

\subsubsection{Asymmetry evolution example}

\begin{figure*}[!ht]
    \centering \includegraphics{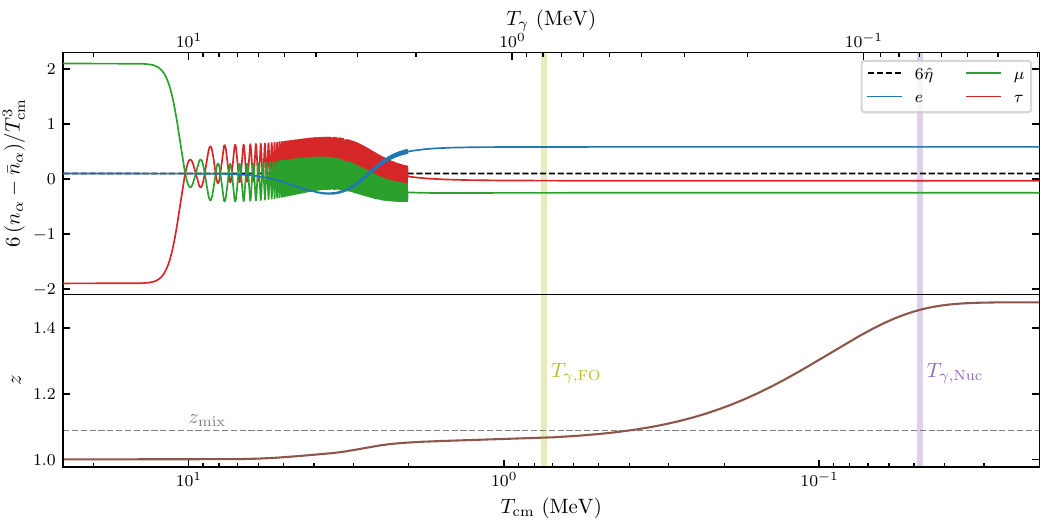}
    \caption{\label{fig:example_evolution} Outputs of neutrino evolution from \nevo, for the initial parameters $\{\xit_\av = 0.1, \, \xit_e = 0.1, \, \xit_\mu - \xit_\tau=4\}$. \emph{Top:} neutrino flavor asymmetries. The frequency of synchronous oscillations increases as $\Tcm$ decreases. \emph{Bottom:} comoving photon temperature. The first increase in $z$, for $10 \, \mathrm{MeV} \geq \Tcm \geq 1 \, \mathrm{MeV}$, is due to the (imperfect) mixing of flavors and the associated increase in entropy, as explained in Sec.~\ref{Sec:equilibration}. The dashed line, denoted $z_\eqb$, would correspond to perfect flavor equilibration, which is not achieved here and highlights the need for an accurate calculation of neutrino evolution. The vertical colored lines identify the approximate temperature of neutron-to-proton weak freeze-out ($T_{\gamma,\mathrm{FO}} \sim \SI{0.8}{\mega \electronvolt}$) and the beginning of deuterium burning ($T_{\gamma,\mathrm{Nuc}} \sim \SI{0.07}{\mega \electronvolt}$).}
\end{figure*}

We draw on Fig.~\ref{fig:example_evolution} an example of result of a \nevo calculation, for the initial parameters $\{\xit_\av = 0.1, \delta \xit_e = 0, \widetilde{\Delta} = 2\}$. The features of asymmetry evolution detailed in~\cite{Froustey2021} are recovered: for $\Tcm \sim \SI{10}{\mega \electronvolt}$, oscillations develop in the $\{\mu,\tau\}$ subspace, before a conversion with the $e$ flavor kicks in for $\Tcm \sim \SI{5}{\mega \electronvolt}$. Throughout the evolution, the average asymmetry is conserved and remains equal to the black dashed line on Fig.~\ref{fig:example_evolution} (top panel). This is a consequence of the structure of the collision terms, which satisfy $\mathrm{Tr}(\mathcal{K}-\overline{\mathcal{K}})=0$. We found that a linearly spaced grid in momenta is the best choice to ensure numerical stability and the conservation of the average asymmetry, contrary to, e.g., the Gauss-Laguerre quadrature used in~\cite{Froustey2020}. Collisions are also responsible for the progressive damping of the amplitude of collective oscillations. At $\Tcm \simeq \SI{2.2}{\mega\electronvolt}$, the solver switches from the ATAO scheme using the full Hamiltonian in Eqs.~\eqref{eq:QKE_rho}--\eqref{eq:QKE_rhobar} (“\ATAOJH” scheme in \cite{Froustey2021}) to effectively average the individual oscillations around $\mathcal{V} \equiv \H_0 + \H_\mathrm{lep}$ only (“\ATAOH” scheme). This is justified by the $\propto \Tcm^{-6}$ increase of the oscillation frequency: these oscillations are so fast that any physical quantity [like the neutron-to-proton ratio set by~\eqref{eq:np_ratio_reac}] is only sensitive to their average, whose evolution is solely dictated by $\mathcal{V}$ and collisions.

On the bottom panel, the dimensionless photon temperature $z = T_\gamma/\Tcm$ increases in two stages: the first “bump” between $\Tcm \sim \SI{10}{\mega \electronvolt}$ and $\Tcm \sim \SI{1}{\mega \electronvolt}$ is due to the overall reduction of the asymmetries because of flavor mixing (we discuss this process in details in Sec.~\ref{Sec:equilibration}). The entropy that was “stored” in the neutrino/antineutrino asymmetry is redistributed among all species through a common increased temperature. A perfect flavor equilibration would lead to the value $z_\eqb$, showed with a dashed grey line. Although it generally slightly overestimates the size of the first increase of $z$, it provides a conservative prediction for the ratio between $T_\gamma$ and $\Tcm$ when the neutron-to-proton interconversion reactions freeze out. As we aim to have an accurate description of neutrino distributions in this period, we adjust the transition temperature $T_\trans$ between the \ATAOJH and \ATAOH schemes by rescaling the default value $\bar{T}_\trans = \SI{2.2}{\mega \electronvolt}$ via $T_\trans = \bar{T}_\trans/z_\eqb$. In addition, there are particular configurations that lead to delayed oscillations, in which case it is crucial to switch to an \ATAOH scheme much later. Our code deals with this issue by decreasing $T_\trans$ if the evolution is particularly “smooth,” which can indicate that there were too few oscillations above $T_\trans$. This adaptive $T_\trans$ is thus in some cases down to $1.0 \, \mathrm{MeV}$ to ensure the accuracy of our results. We demonstrate in Appendix~\ref{App:Ttrans} the need for this method to probe the actual features of asymmetry mixing.

The second “bump” in $z$, for $\Tcm \sim \SI{0.1}{\mega\electronvolt}$, is due to electron-positron annihilations, which mainly reheat photons and only marginally neutrinos. During this phase, the asymmetries $\eta_\alpha$ are constant (see top panel); we note their values $\eta_\alpha^\mathrm{BBN}$. We identify the temperatures corresponding to the weak freeze-out of $n/p$ conversion ($T_{\gamma,\mathrm{FO}} \sim \SI{0.8}{\mega \electronvolt}$) and the beginning of nucleosynthesis ($T_{\gamma,\mathrm{Nuc}} \sim \SI{0.07}{\mega \electronvolt}$). The value of the electron neutrino asymmetry at $T_{\gamma,\mathrm{FO}}$ is key to determine the neutron-to-proton ratio at the onset of BBN and then the final abundances~\cite{Froustey2019}. Note that, due to the different values of $z$ at these two epochs, the ratios $T_{\gamma, \mathrm{FO}}/T_\mathrm{cm,FO}$ and $T_{\gamma,\mathrm{Nuc}}/T_\mathrm{cm,Nuc}$ are different. 

\subsection{Big Bang Nucleosynthesis}
\label{subsec:BBN}

The output from \nevo, namely, $\Neff$ and (anti)-neutrino distributions (thus including the spectral distortions incurred in the decoupling process), is used as input in the BBN code \primat~\cite{Pitrou_2018PhysRept}. This \emph{Mathematica} code computes the primordial abundances of light elements up to CNO isotopes in three main steps:
\begin{enumerate}[label=(\roman*)]
    \item first, the background thermodynamics are determined [i.e., the function $a(T_\gamma)$ and inversely] consistently with \nevo results — neglecting the effect of baryonic matter is justified by the very small value of the baryon-to-photon ratio $\eta_\mathrm{b} \simeq \num{6e-10}$;
    \item second, the weak neutron-to-proton interconversion rates are tabulated as a function of temperature, including radiative, finite nucleon mass, and weak magnetism corrections and also taking into account the non-thermal part of electron (anti)neutrino distribution functions from \nevo ;
    \item finally, the network of nuclear reactions along with weak interactions interconverting neutrons and protons is numerically solved down to a temperature of $\SI{6e7}{\kelvin}$.
\end{enumerate}

BBN abundances depend on the baryon density parameter $\omega_b \equiv \Omega_b h^2$, with $\Omega_b$ the baryonic fraction of the critical density today, and $h = H_0/{100 \, \mathrm{km \, s^{-1} \, Mpc^{-1}}}$ the present Hubble parameter. $\omega_b$ is an input parameter in \primat. Given the tight constraints obtained by Planck, which correspond to a variation range allowed for $\omega_b$ of less than \SI{1}{\percent}, we only use the central value from~\cite{Planck:2018vyg}, $\omega_b^\primat = 0.02242$, in the full \primat calculation, and estimate the abundances for $\omega_b \neq \omega_b^\primat$ assuming a linear variation.

\paragraph*{Neutrinos and BBN —} Cosmological neutrinos influence BBN in two ways. The principal effect corresponds to the change of the neutron-to-proton ratio, which is set by the weak reactions:
\begin{equation}
    \label{eq:np_ratio_reac}
    \begin{aligned}
        n + \nu_e &\longleftrightarrow p + e^- \, , \\
        n &\longleftrightarrow p + e^- + \bar{\nu}_e \, , \\
        n + e^+ &\longleftrightarrow p + \bar{\nu}_e \, .
    \end{aligned}
\end{equation}
If we assume that neutrinos are at equilibrium when the above reactions freeze out (which happens for $T_{\gamma,\mathrm{FO}} \sim \SI{0.8}{\mega \electronvolt}$), with degeneracy parameters $\xi_\alpha^\mathrm{BBN}$, the electron (anti)neutrino chemical potential modifies the equilibrium value via: 
\begin{equation}
\label{eq:npratio}
    \frac{n_n}{n_p} = \left.\frac{n_n}{n_p}\right\rvert_{\xi_e = 0} \times e^{-\xi_e^\mathrm{BBN}} \, .
\end{equation}
In addition, the increased energy density compared to the $\xi = 0$ situation leads to a contribution to $\Neff$:
\begin{equation}
\label{eq:dNeff_xi}
    \Delta \Neff = \sum_{\alpha = e, \mu, \tau}{\left[\frac{30}{7} \left(\frac{\xi_\alpha^\mathrm{BBN}}{\pi}\right)^2 + \frac{15}{7} \left(\frac{\xi_\alpha^\mathrm{BBN}}{\pi}\right)^4 \right]} \, .
\end{equation}
This formula does not take into account a possible reheating $z_\alpha^\mathrm{BBN} > 1$, a valid assumption if we assume that all initial asymmetries are equal, which is virtually the only case when this formula can be applied anyway. For a given photon temperature, the increased energy density leads to a higher expansion rate, which reduces the time left to neutrinos to undergo beta decay between weak freeze-out ($T_{\gamma, \mathrm{FO}}$) and the onset of nucleosynthesis ($T_{\gamma, \mathrm{Nuc}} \sim \SI{0.07}{\mega \electronvolt}$). As a consequence, this so-called “\emph{clock effect}”~\cite{Froustey2019,Dodelson:1992km,Fields:1992zb} leads to a higher neutron-to-proton ratio. However, since it scales like $\propto (\xi_\alpha^\mathrm{BBN})^2$, this effect is very small and is hidden by the changes given in Eq.~\eqref{eq:npratio}. Note finally that, because of the same clock effect, there is less time to destroy deuterium during BBN, which results in a slight increase in $\DH$~\cite{Froustey2019}.

The net effect on helium-4 and deuterium abundances coming from an electron (anti)neutrino chemical potential during BBN was estimated numerically in~\cite{Pitrou_2018PhysRept}:
\begin{equation}
    \label{eq:YpDHxi_old}
    \frac{\Yp}{\Yp\rvert_{\xi_e = 0}} \simeq e^{- 0.96 \xi_e^\mathrm{BBN}} \ , \quad \frac{\DH}{\DH\rvert_{\xi_e = 0}} \simeq e^{- 0.53 \xi_e^\mathrm{BBN}} \, .
\end{equation}
In the general case, a quantity such as $\xi_e^\mathrm{BBN}$ is ill-defined, notably because neutrinos are not at equilibrium, and because neutrino and photon temperatures differ, which is not assumed in Eq.~\eqref{eq:npratio}. However, since neutron-to-proton interconversion freeze-out occurs before the reheating due to $e^- e^+$ annihilations (see Fig.~\ref{fig:example_evolution}), we can estimate an effective $\xi_e^\mathrm{BBN}$ parameter as:
\begin{align}
\xi_e^\mathrm{BBN} \sim \left[ \frac{6(n_e - \bar{n}_e)}{T_{\gamma}^3}\right]_\mathrm{FO} &= \left[\frac{6(n_e - \bar{n}_e)}{z^3 \Tcm^3}\right]_\mathrm{FO} \nonumber \\
&= 6 \left[\frac{\eta_e}{z^3}\right]_\mathrm{FO} \\
&= 6 \left[\frac{\eta_e}{z^3}\right]_\text{final} \times \left(\frac{z_\text{final}}{z_\mathrm{FO}}\right)^3 \, , \nonumber
\end{align}
where we used that the comoving asymmetry $6(n_e - \bar{n}_e)/\Tcm^3$ is constant during the entire BBN epoch (see Fig.~\ref{fig:example_evolution}). The ratio of photon temperatures between weak freeze-out and after BBN is mostly independent of the particular asymmetries: it is set by the electron/positron annihilations and is given up to a few percents by $z_\text{final}/z_\mathrm{FO} \simeq (11/4)^{1/3}$~\cite{KolbTurner}. Therefore, this constant prefactor put aside, the effect on primordial abundances should be mostly determined by the parameter $[\eta_e/z^3]_\text{final}$, as will be confirmed in Sec.~\ref{section:constraints}.

Since helium-4 and deuterium are the only light elements for which the primordial abundances are well-enough experimentally constrained, the output of our calculation is the set of quantities $\{\Neff,\Yp,\DH\}$ that can be compared to experimental values for $\Yp$ and $\DH$, and to CMB experiment posteriors for $(\Neff,\Yp)$. 

\subsection{Likelihood}
\label{subsec:likelihood}

As explained above, the input parameters of our calculation is the set of initial asymmetries $\{\xi_\alpha, \alpha=e,\mu,\tau\}$ and the baryon abundance $\omega_b$. When combining CMB and BBN experimental constraints, we get the likelihood [see Eq.~\eqref{eq:app_likelihood_tomarg} and more generally Appendix~\ref{App:statistics}] 
\begin{multline}
\label{eq:likelihood_tomarg}
\L(\xi_\alpha, \omega_b \rvert C_\ell^\obs,\Yp^\obs,\DH^\obs) \\
= \L_\mathrm{CMB} \left(\omega_b, \Neffxi^\code, {\Yp}^\code_{(\xi_\alpha,\omega_b)} \right) \\
\times \mathcal{N}\left(\Yp^\obs ; {\Yp}^\code_{(\xi_\alpha,\omega_b)}, \sigma_{\Yp}\right) \\ \times \mathcal{N}\left(\DH^\obs ; {\DH}^\code_{(\xi_\alpha,\omega_b)}, \sigma_{\DH}\right) \, .
\end{multline}
We marginalize over the baryon density, such that the final likelihood is given by
\begin{equation}
\label{eq:likelihood_final}
    \L(\xi_\alpha) = \int{\dd{\omega_b} \, \L(\xi_\alpha, \omega_b \rvert C_\ell^\obs,\Yp^\obs,\DH^\obs)} \, .
\end{equation}
Since the allowed space parameter for $\omega_b$ is very narrow [see Eqs.~\eqref{eq:mean_omb_Planck} and \eqref{eq:mean_omb_allCMB}], we only use the central value in \primat and linearly extrapolate the obtained abundances, that is
\begin{equation}
    {\Yp}^\code_{(\xi_\alpha, \omega_b)} \simeq {\Yp}^\code_{(\xi_\alpha,\omega_b^\primat)} + (\omega_b - \omega_b^\primat) \left. \frac{\dd \Yp^\code}{\dd \omega_b} \right\rvert_{(\xi_\alpha, \omega_b^\primat)} ,
\end{equation}
the derivative on the right-hand side being numerically evaluated in \primat for each calculation.

\subsubsection{BBN abundances}

Our reference value for the helium-4 abundance is the one obtained by Aver et al.~\cite{Aver2020}, while we use the recommended deuterium abundance value from Kislitsyn et al.~\cite{Kislitsyn:2024jvk} (see also~\cite{Cooke:2017cwo,Guarneri:2024qxi}), which is slightly smaller but consistent with the PDG-recommended value~\cite{PDG-2022} (as they use a different selection of available measurements). The values read:
\begin{align}
    \Yp^\obs &= 0.2453 \pm 0.0034 &&\text{\protect\cite{Aver2020}} \, , \label{eq:Yp_mes} \\
    \DH^\obs &= (2.533 \pm 0.024) \times 10^{-5} &&\text{\protect\cite{Kislitsyn:2024jvk}} \, . \label{eq:DH_mes}
\end{align}
Recently, the EMPRESS survey~\cite{Matsumoto:2022tlr} of extremely metal-poor galaxies obtained a smaller value for $\Yp^\obs$ with a $\sim 1 \sigma$ difference compared to \eqref{eq:Yp_mes}, namely:
\begin{equation}
\label{eq:Yp_empress}
    \Yp^\obs\rvert_\mathrm{EMPRESS} = \num{0.2370}^{+ \num{0.0034}}_{-\num{0.0033}} \qquad \text{\protect\cite{Matsumoto:2022tlr}} \, .
\end{equation}
This value would indicate a non-zero lepton asymmetry at the epoch of BBN~\cite{Burns:2022hkq,Escudero:2022okz}. We will analyze our results using either \eqref{eq:Yp_mes} or \eqref{eq:Yp_empress} in the following.

The deuterium abundance obtained with \primat for zero asymmetries is in slight tension with the value~\eqref{eq:DH_mes}~\cite{Pitrou2020}. This is a feature of \primat, as other codes predict, on the contrary, a value for $\DH$ in agreement with \eqref{eq:DH_mes}~\cite{Parthenope_revolutions,Yeh2020}. The difference comes from different selections of measurements of nuclear data, which result in different nuclear rates for the $\mathrm{D(d,n)^{3}He}$ and $\mathrm{D(d,p)^{3}He}$ reactions. Nevertheless, our results are not qualitatively modified by this feature — see Appendix~\ref{App:deuterium} for the same analysis as Sec.~\ref{section:constraints} but removing the $\DH$ measurement from our likelihood. 

\subsubsection{CMB anisotropies}

We approximate the likelihood obtained from CMB anisotropies by a normal distribution in the three parameters $(\omega_b, \Neff, \Yp)$, see Eq.~\eqref{eq:likelihood_CMB}. We must therefore provide the means of the former parameters and their associated covariance matrix. We use the Markov chain Monte Carlo (MCMC) sampler \texttt{Cobaya}~\cite{Torrado:2020dgo} to estimate the means and covariance matrices from CMB experiments combined with all available BAO data~\cite{Beutler:2011hx,Ross:2014qpa,BOSS:2016wmc,eBOSS:2020yzd}. For Planck~\cite{Planck:2018vyg,Planck:2019nip,Planck18res}~+ BAO, we find the preferred values (denoted  with “Planck” in the following)
\begin{subequations}
\label{eq:means_Planck}
\begin{align}
    \omega_b\rvert_\text{Planck} &= 0.02237 \pm 0.00018 \, , \label{eq:mean_omb_Planck} \\
    \Neff\rvert_\text{Planck} &= \num{3.02} \pm \num{0.28} \, , \\
    \Yp\rvert_\text{Planck} &= \num{0.244} \pm \num{0.018} \, ,
\end{align}
\end{subequations}
and the covariance matrix
\begin{equation}
\label{eq:cov_planck}
\Sigma_\text{Planck} = \begin{pmatrix}
    \num{3.25e-8} & \num{9.88e-6} & \num{1.03e-6} \\
    \num{9.88e-6} & \num{0.0794} & - 0.00378 \\
    \num{1.03e-6} & -0.00378 & 0.000342
\end{pmatrix} \, .
\end{equation}
We also considered the combination of all most recent CMB experiments, adding BICEP/Keck~\cite{BICEP:2021xfz}, ACT~\cite{ACT:2020frw,ACT:2020gnv,Carron:2022eyg,ACT:2023dou,ACT:2023kun} and SPT-3G~\cite{SPT-3G:2022hvq} to Planck+BAO. The means and the covariance matrix, denoted “all-CMB” in the following, read:
\begin{subequations}
\label{eq:means_allCMB}
\begin{align}
    \omega_b\rvert_\allCMB &= 0.02217 \pm 0.00016 \, , \label{eq:mean_omb_allCMB} \\
    \Neff\rvert_\allCMB &= \num{2.99} \pm \num{0.24} \, , \\
    \Yp\rvert_\allCMB &= \num{0.235} \pm \num{0.015} \, , \label{eq:mean_Yp_allCMB}
\end{align}
\end{subequations}
and
\begin{equation}
\label{eq:cov_allCMB}
\Sigma_\allCMB = \begin{pmatrix}
    \num{2.46e-8} & \num{8.61e-6} & \num{6.65e-7} \\
    \num{8.61e-6} & \num{0.0592} & - 0.00283 \\
    \num{6.65e-7} & -0.00283 & 0.000234
\end{pmatrix} \, .
\end{equation}
Note that the primordial helium abundance is currently much less constrained by CMB experiments compared to the spectroscopic measurements [Eqs.~\eqref{eq:Yp_mes} and~\eqref{eq:Yp_empress}], which is expected since the effect of $\Yp$ on CMB is tenuous (it affects the damping tail of the CMB anisotropies by modifying the density of free electrons between helium and hydrogen recombination~\cite{Planck:2015fie}).

\section{Asymmetry equilibration and reheating}
\label{Sec:equilibration}

Initially large asymmetries do not necessarily lead to high values of $\Neff$ or to large asymmetries at the epoch of BBN. In particular, the “equilibration” of asymmetries is associated to a global reheating of all the species that are coupled when this equilibration takes place. As perfect flavor equilibration of asymmetries is the standard assumption in the literature, we discuss first the consequences of this ideal situation, which provides a useful estimate of the evolution of $T_\gamma/\Tcm$.

\subsection{Analytical description}

\subsubsection{Conservation laws}

If neutrino flavors could fully equilibrate their asymmetries while being fully coupled to the electromagnetic plasma, their distributions after this equilibration would entirely be characterized by the common temperature $z_\eqb$ and the common reduced chemical potential $\xi_\eqb$. Finding two conserved quantities is sufficient to determine their values. 

The first conservation law is exact and is provided by the conservation of $\hat \eta$ defined in Eq.~\eqref{Eq:defetabar}. Indeed, we have for a given flavor~\cite{Froustey2021}
\begin{equation}
    \frac{\dd \eta_\alpha}{\dd x} = \int{\frac{y^2 \dd{y}}{2 \pi^2}\left(- \ii \left[\H_0 + \H_\mathrm{lep},\vrho+\bvrho\right]_{\alpha \alpha} + \mathcal{K}_{\alpha \alpha} -\overline{\mathcal{K}}_{\alpha \alpha}\right)} \, ,
\end{equation}
such that $3 \dd \hat{\eta}/\dd x = \sum_{\alpha}{\dd\eta_\alpha/\dd x}=0$, since the trace of a commutator is zero and $\mathrm{Tr}(\mathcal{K} - \overline{\mathcal{K}}) = 0$.

The second conservation equation corresponds to the energy density. In the approximation that electrons and positrons are relativistic, that is for temperatures such that $x/z \ll 1$, all the relevant energy densities redshift as $\Tcm^4$. In that case, $\hat \rho \equiv \rho_\mathrm{tot}/\Tcm^4$, where $\rho_\mathrm{tot}$ is the sum on all energy densities, is a conserved quantity. For a plasma made of three (anti)neutrino flavors, photons, electrons, and positrons with a shared reduced temperature $z=z_{e,\mu,\tau}$, the total comoving energy density is given by (see Appendix~\ref{App:thermo})
\begin{equation}
\hat \rho(z,\{ \xi_\alpha \})= z^4\left[\frac{43 \pi^2}{120} + \frac{1}{4}\sum_\alpha \left(\xi_\alpha^2 + \frac{\xi_\alpha^4}{2 \pi^2}\right)\right] \, .
\end{equation}
If there is a full equilibration in the regime where electrons and positrons are still relativistic, we can use these conservation laws to obtain the shared temperature $z_\eqb$ and the shared neutrino chemical potential $\xi_\eqb$, via
\begin{equation}
\begin{aligned}
\hat \eta(\{z_\alpha=1,\xi_\alpha\}) &= \hat \eta(\{z_\eqb,\xi_\eqb\}) \, , \\
\hat \rho(z=1, \{ \xi_\alpha \} ) &= \hat \rho(z_\eqb, \{ \xi_\eqb \} ) \, ,
\end{aligned}
\end{equation}
that is,
\begin{subequations}
\label{eq:zmix_ximix}
\begin{align}
\frac{1}{3} \sum_{\alpha}{\xit_\alpha} &= z_\eqb^3 \tilde \xi_\eqb \, , \\
\frac{43 \pi^2}{120} + \frac{1}{4}\sum_\alpha \left(\xi_\alpha^2 + \frac{\xi_\alpha^4}{2 \pi^2}\right) &=  \frac{43 \pi^2}{120}z_\eqb^4  \\ &+ \frac{3}{4}\left(\xi_\eqb^2 + \frac{\xi_\eqb^4}{2 \pi^2}\right)z_\eqb^4 \, . \nonumber
\end{align}
\end{subequations}

\subsubsection{Entropy variation}

This full equilibration is a nonreversible process as the total comoving entropy is not conserved. Indeed, from the thermodynamic identity
\begin{equation}
\frac{\dd}{\dd T} \left(\frac{\rho + P}{T} \right) = \frac{1}{T} \frac{\dd \rho}{\dd T} + n \frac{\dd}{\dd T} \left(\frac{\mu}{T}\right) \, ,
\end{equation}
valid for each species individually, together with $\rho+P = 4/3 \rho$ and the conservation of $\hat{\rho}$ when electrons and positrons are assumed to be fully relativistic, we obtain (using $\bar \mu_\alpha = - \mu_\alpha$ to relate neutrino and antineutrino chemical potentials)
\begin{equation}\label{Eq:evolentropy}
\frac{\dd }{\dd x}\left(s_\mathrm{tot} / \Tcm^3\right) = - \sum_\alpha \xi_\alpha \frac{\dd \eta_\alpha}{\dd x} \,.
\end{equation}
Alternatively, this can be obtained from a direct computation (see Appendix \ref{App:thermo} for the useful formulae), with
\begin{equation}
\frac{s_\mathrm{tot}}{\Tcm^3} = z^3 \left(\frac{43 \pi^2}{90}  + \frac{1}{6}\sum_\alpha \xi_\alpha^2 \right)\,,
\end{equation}
and using the constancy of $\hat{\rho}$. 

Since $\dd \hat{\eta} / \dd x = 0$, we can rewrite \eqref{Eq:evolentropy} as
\begin{equation}
\label{eq:evolentropy_2}
    \frac{\dd}{\dd x} \left(s_\mathrm{tot} / \Tcm^3 \right) = - \sum_{\alpha}{(\xi_\alpha - \xi_\av) \frac{\dd}{\dd x}(\eta_\alpha - \hat{\eta})} \, ,
\end{equation}
with $\xi_\av = (\sum_\alpha \xi_\alpha)/3$. If a given flavor $\alpha$ has a larger (resp.~smaller) asymmetry than the average, that is $\xi_\alpha - \xi_\av > 0$ (resp. $<0$), the “equilibration” leads to\footnote{$\xi_\alpha \longmapsto \eta_\alpha(\xi_\alpha)$ is a monotonically increasing function.} a decrease (resp. increase) of $\eta_\alpha - \hat{\eta}$. Therefore, the right-hand side of~\eqref{eq:evolentropy_2} is positive, which corresponds to an increase of the total comoving entropy density.

\subsubsection{Special cases}

If the initial asymmetries are small, that is for $\xi_\alpha \simeq \tilde \xi_\alpha \ll 1$, then keeping only terms up to order $\xi_\alpha^2$, the postequilibration plasma is characterized by the following solution of Eq.~\eqref{eq:zmix_ximix}:
\begin{subequations}
\begin{align}
\xi_\eqb &\simeq \frac{1}{3} \sum_\alpha \xi_\alpha\,,\\
z_\eqb &\simeq 1 + \frac{15}{86 \pi^2}\left(\sum_\alpha\xi_\alpha^2-3 \xi_\eqb^2\right)\,.
\end{align}
\end{subequations}
The reheating of the plasma, given by $z_\eqb-1$, is therefore very small since it is at least quadratic in the $\xi_\alpha$. The final chemical potential of such a mildly asymmetric neutrino plasma is simply the average of the initial chemical potentials. The variation of entropy in that case is 
\begin{equation}
\Delta \left(s_\mathrm{tot} / \Tcm^3\right) = \frac{1}{4}\left(\frac{1}{3}\sum_\alpha \xi_\alpha^2  - \xi_\eqb^2\right)\,,
\end{equation}
which vanishes as expected if all initial potentials are equal ($\xi_\alpha = \xi_\eqb$).

Another interesting case is when chemical potentials are large, that is  when $\xi_\alpha \gg 1 $. If the different $\xi_\alpha$ are large but have similar values, equilibration also tends to the average chemical potential and no substantial reheating of the plasma takes place. However, an interesting situation arises when the initial potentials are large, but with typically very different values such that the average initial asymmetry is small compared to the energy excess. This situation is typically realized when 
\begin{equation}\label{Eq:xi443xi3}
\left(\sum_\alpha \xi_\alpha^4\right)^{3/4} \gg \sum_\alpha \xi_\alpha^3\,.
\end{equation}
The final asymmetry when full equilibration has occurred is then very small and given by
\begin{subequations}
\begin{align}
\xi_\eqb \simeq \tilde \xi_\eqb &= \frac{1}{3 z_\eqb^3} \sum_\alpha \tilde \xi_\alpha = \frac{\xit_\av}{z_\eqb^3} \ll 1 \label{eq:ximix_2}\\
z_\eqb^4 &\simeq \frac{15}{43 \pi ^4} \sum_\alpha \xi_\alpha^4\,. \label{eq:zmix_2}
\end{align}
\end{subequations}
Note that this occurs even if $\sum_\alpha \xi_\alpha^3$ is not small, as long as the condition \eqref{Eq:xi443xi3} is satisfied. In that case, the full equilibration can only be realized by a strong energy exchange with the electromagnetic plasma which receives a substantial part of the energy excess contained in the initially very asymmetric neutrino distributions. This implies that $z_\eqb$ substantially departs from unity and the final asymmetry $\tilde \xi_\eqb$ ends up being smaller than unity. Consequently, $\Neff$ does not depart much from its standard value without initial asymmetries. As expected, this is associated with an important entropy variation
\begin{equation}
\Delta \left(s_\mathrm{tot} / \Tcm^3\right) \simeq \frac{1}{6 \pi}\left(\frac{43}{15}\right)^{1/4}\left(\sum_\alpha \xi_\alpha^4\right)^{3/4}\,.
\end{equation}

\subsection{Numerical examples}

\begin{figure}[!ht]
    \centering \includegraphics{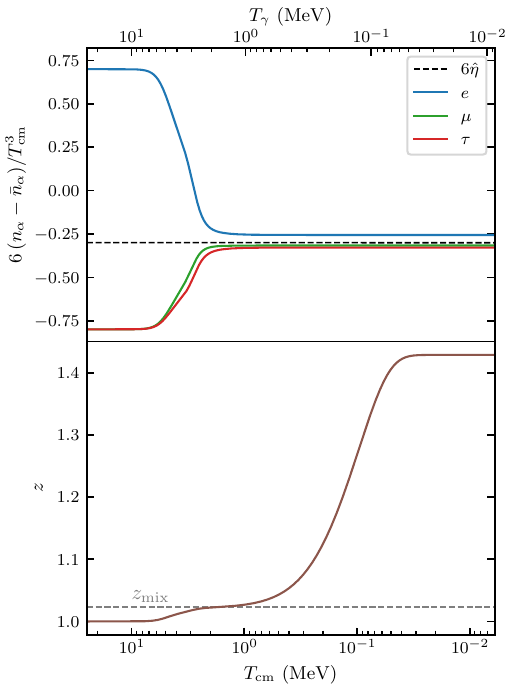}
    \caption{\label{fig:example_evolution_fullmix} Evolution of neutrino asymmetries (\emph{Top}) and comoving photon temperature (\emph{Bottom}) from \nevo for the initial parameters $\{\xit_\av = -0.3, \xit_e = 0.7, \tilde{\Delta}=0\}$. The equilibration of flavor asymmetries being almost total, the amount of reheating between $\SI{10}{\mega \electronvolt}$ and $\SI{1}{\mega \electronvolt}$ is very well described by $z_\eqb$.}
\end{figure}

The previous analytical estimates provide a good description of the evolution of the asymmetries and the photon temperature when equilibration is complete. Such an example is represented Fig.~\ref{fig:example_evolution_fullmix}, which corresponds to the initial parameters $\{\xit_\av = -0.3, \delta \xit_e = 1.0, \Deltat = 0\}$. One must not be led to believe that $\Deltat = 0$, which is the case on Fig.~\ref{fig:example_evolution_fullmix} and not on Fig.~\ref{fig:example_evolution}, is a criterion for flavor equilibration. This is just a coincidence, and results presented hereafter (see, e.g., Figs.~\ref{fig:etae_xie} and \ref{fig:etae_delta}) will highlight that (in)complete equilibration is a highly non-linear problem for which there is no general criterion.

\begin{figure}[!ht]
    \centering
    \includegraphics{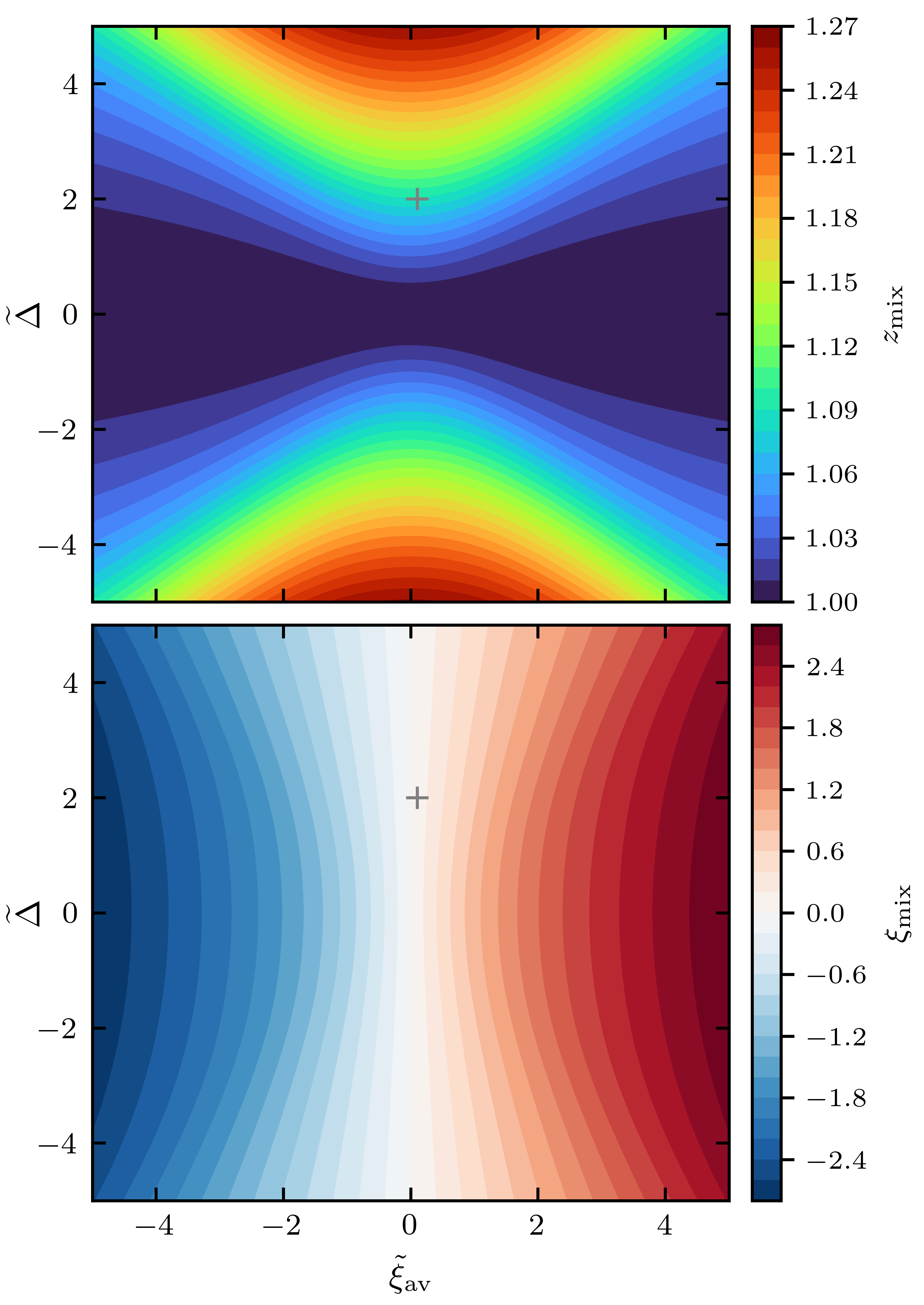}
    \caption{\label{fig:zmix_ximix} Dimensionless photon temperature $z_\eqb$ (\emph{Top}) and common degeneracy parameter $\xi_\eqb$ (\emph{Bottom}) after perfect flavor equilibration when electrons and positrons are still relativistic, given by the solution of Eq.~\eqref{eq:zmix_ximix}. The example point shown on Fig.~\ref{fig:example_evolution} corresponds to the gray cross, with coordinates $(0.1,2)$.}
\end{figure}

We represent on Fig.~\ref{fig:zmix_ximix} the “fully mixed” parameters $(z_\eqb,\xi_\eqb)$ for a range of initial conditions $(\xit_\av, \delta \xit_e = 0, \Deltat)$. The vertical band around $\xit_\av \simeq 0$ corresponds to the second situation discussed above [Eq.~\eqref{Eq:xi443xi3}]: when varying $\Deltat$, we see on the bottom panel that $\xi_\eqb \simeq 0$, in agreement with Eq.~\eqref{eq:ximix_2}, while on the top panel we see the rapid dependence of $z_\eqb$ with $\xi_\alpha$, consistently with Eq.~\eqref{eq:zmix_2}.

Let us stress again that these limiting results are only valid when the equilibration between flavor species is complete, and takes place before electrons and positrons become nonrelativistic. This is not the case in general: equilibration can be incomplete (see for instance Fig.~\ref{fig:example_evolution}), or collective oscillations can be delayed (see for instance Sec.~\ref{subsec:EBO}). Therefore, in general, computing correctly neutrino flavor evolution is crucial to adequately constrain the primordial asymmetries, as we shall show in the next section. Note that, as explained in Sec.~\ref{subsec:Nu_transport}, we use $z_\eqb$ in the general case to provide an approximate rescaling between $\Tcm$ and $T_\gamma$ during the epoch prior to the weak freeze-out of $n \leftrightarrow p$ reactions. This allows us to always switch from \ATAOJH to \ATAOH schemes in the same range of $T_\gamma$.

\section{Constraints on primordial neutrino asymmetries}
\label{section:constraints}

A full exploration of the three-dimensional space of initial asymmetries $(\xi_e,\xi_\mu,\xi_\tau)$ is out of reach for computational reasons. We thus restrict the parameter range, first assuming equal asymmetries (the usual assumption made in the literature), then generalizing to two-dimensional extensions. 

Our various datasets are available on \emph{Zenodo}~\cite{Data_Zenodo}.


\subsection{Equal asymmetries}
\label{subsec:equal_asymmetries}

Based on the generic trend that flavor oscillations tend to “equilibrate” asymmetries between the different flavors, previous studies have often assumed that, throughout the range of temperatures of interest, flavor asymmetries were equal. With such an assumption, there is a priori no need for a full dynamical calculation and one can directly include a neutrino chemical potential in a BBN code to assess the change in primordial abundances and thus constrain $\xi$. Note, however, that this neglects the small out-of-equilibrium effects that can only be tracked with a dynamical calculation and are present in our results.

As a check that our general calculation provides consistent results with the existing literature, we first present the likelihood obtained by assuming, in the initial conditions of \nevo, $\xi_e = \xi_\mu = \xi_\tau$, which corresponds to the parameters $\delta \xit_e = 0$ and $\Deltat = 0$. 

\begin{figure}[!ht]
    \centering
    \includegraphics{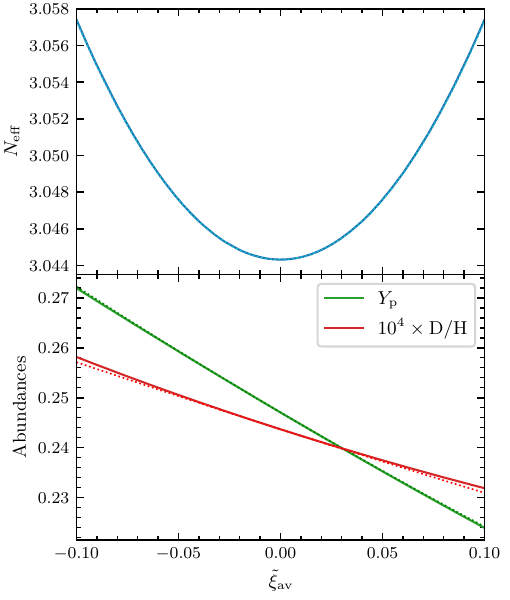}
    \caption{\label{fig:output_equalxi} Output from neutrino evolution and BBN calculation as a function of the initial degeneracy parameter, equal for all three neutrino species. \emph{Top:} $\Neff$, which varies by \SI{0.4}{\percent} over the range of asymmetries explored. \emph{Bottom:} primordial abundances of helium-4 and deuterium (calculated for the fiducial baryon fraction), which follow closely the dependency~\eqref{eq:YpDHxi}. Note that the deuterium abundance is multiplied by $10^4$ to be plotted in the same range as $\Yp$.}
\end{figure}

\begin{figure}[!ht]
    \centering
    \includegraphics{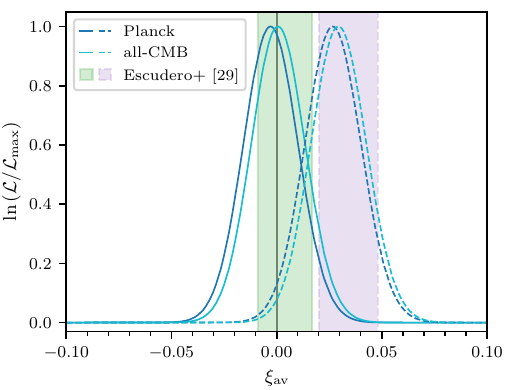}
    \caption{\label{fig:likelihood_equalxi} Normalized likelihood as a function of the initial reduced chemical potential, equal for all three neutrino species. Solid lines correspond to $\Yp\rvert_\mathrm{Aver}$ [Eq.~\eqref{eq:Yp_mes}], dashed lines to $\Yp\rvert_\mathrm{EMPRESS}$ [Eq.~\eqref{eq:Yp_empress}]. The colored bands correspond to the \SI{68}{\percent} intervals obtained in \cite{Escudero:2022okz}, with $\Yp\rvert_\mathrm{Aver}$ (green) and $\Yp\rvert_\mathrm{EMPRESS}$ (purple), see Eq.~\eqref{eq:Escudero_xi}.}
\end{figure}

The output values of $\Neff$ and the primordial abundances $\Yp$ and $\DH$ (calculated for the fiducial baryon abundance $\omega_b^\primat = 0.02242$) are shown on Fig.~\ref{fig:output_equalxi}. The associated likelihood, introduced in Sec.~\ref{subsec:likelihood}, is represented on Fig.~\ref{fig:likelihood_equalxi}. Note that we restrict our study to the range $\xit_\av \in [-0.1,0.1]$, for which $\xit_\av \simeq \xi_\av$ up to a relative difference of $10^{-3}$. The expected dependency of $\Neff$ on $\xi_\av$ is given by Eq.~\eqref{eq:dNeff_xi} with $\xi_e = \xi_\mu = \xi_\tau = \xi_\av$:
\begin{equation}
\label{eq:dNeff_xi_equal}
    \Neff(\{\xi_\alpha = \xi_\av\}) = \Neff(0) + \frac{90}{7} \left(\frac{\xi_\av}{\pi}\right)^2 + \frac{45}{7} \left(\frac{\xi_\av}{\pi}\right)^4 \, .
\end{equation}
This expression is drawn as a light blue dotted line on the top panel of Fig.~\ref{fig:output_equalxi}, which superimposes very well with the numerical result (we emphasize however that our results include the subdominant effects of the spectral distortions inherited from neutrino decoupling). Concerning the abundances, the dotted lines on the bottom panel are fits in the form~\eqref{eq:YpDHxi_old}, which read:
\begin{equation}
    \label{eq:YpDHxi}
    \frac{\Yp}{\Yp\rvert_{\xi_e = 0}} \simeq e^{- 0.97 \xi_e^\mathrm{BBN}} \ , \quad \frac{\DH}{\DH\rvert_{\xi_e = 0}} \simeq e^{- 0.54 \xi_e^\mathrm{BBN}} \, .
\end{equation}
The small coefficient differences between Eqs.~\eqref{eq:YpDHxi_old} and \eqref{eq:YpDHxi} can be attributed to the several updates of \primat since the publication of~\cite{Pitrou_2018PhysRept}.

Our results are consistent with the existing literature. In particular, we represent on Fig.~\ref{fig:likelihood_equalxi} the recent constraints obtained by Escudero et al.~\cite{Escudero:2022okz} assuming equal asymmetries. They perform different analyses by inputting directly neutrino chemical potentials in a BBN calculation for different choices of nuclear rates (following \primat or \parthenope) and fixing $\Neff = 3.044$ or letting it be a free parameter that is constrained concurrently. Our calculation is different in that regard, as $\Neff$ is not an external, additional parameter, but it is a consequence of neutrino decoupling for a given value of $\xi_\av$. However, given the range of values taken by $\Neff$ (\SI{0.4}{\percent} variation, see Fig.~\ref{fig:output_equalxi}), it is reasonable to compare our constraints with the $\Neff = 3.044$ case in \cite{Escudero:2022okz}. Therefore, the comparable values are:
\begin{equation}
\label{eq:Escudero_xi}
    \begin{aligned}
        \xi_\av^{\mathrm{[Escudero+]}} &= 0.004 \pm 0.013 &&(\text{\cite{Escudero:2022okz}}, \, \Yp\rvert_\mathrm{Aver}) \, , \\
        \xi_\av^{\mathrm{[Escudero+]}} &= 0.034 \pm 0.014 &&(\text{\cite{Escudero:2022okz}}, \, \Yp\rvert_\mathrm{EMPRESS}) \, .
    \end{aligned}
\end{equation}
These confidence intervals are represented in colored bands on Fig.~\ref{fig:likelihood_equalxi}. Our results are in good agreement, noting that there are a few differences in our implementations: {(i)} we take into account nonthermal effects associated to neutrino decoupling, which are enhanced for a nonzero lepton asymmetry~\cite{Grohs:2016cuu}; {(ii)} we use a slightly different value of the deuterium abundance, viz., \eqref{eq:DH_mes} instead of the value from~\cite{PDG-2022}; {(iii)} we include more recent BAO data in our MCMC analysis of CMB experiments. All this leads to small differences with~\cite{Escudero:2022okz}. Although assuming equal asymmetries is a very limiting assumption that artificially restricts the allowed parameter space, we quote, for future reference, our \SI{68}{\percent} confidence intervals for $\xi_\av$, using all current CMB experimental data:
\begin{equation}
\label{eq:Equal_xi}
    \begin{aligned}
        \xi_\av &= 0.001 \pm 0.013 &&( \Yp\rvert_\mathrm{Aver}) \, , \\
        \xi_\av &= 0.030 \pm 0.013 &&( \Yp\rvert_\mathrm{EMPRESS}) \, .
    \end{aligned}
\end{equation}
Therefore, the EMPRESS measurement is indeed consistent with a nonzero lepton asymmetry at BBN, but given the assumptions that lead to this result, one should more precisely say that \emph{if all neutrino asymmetries are supposed equal, the EMPRESS measurement favors a nonzero value for this common lepton asymmetry.}

\begin{figure}[!ht]
    \centering
    \includegraphics{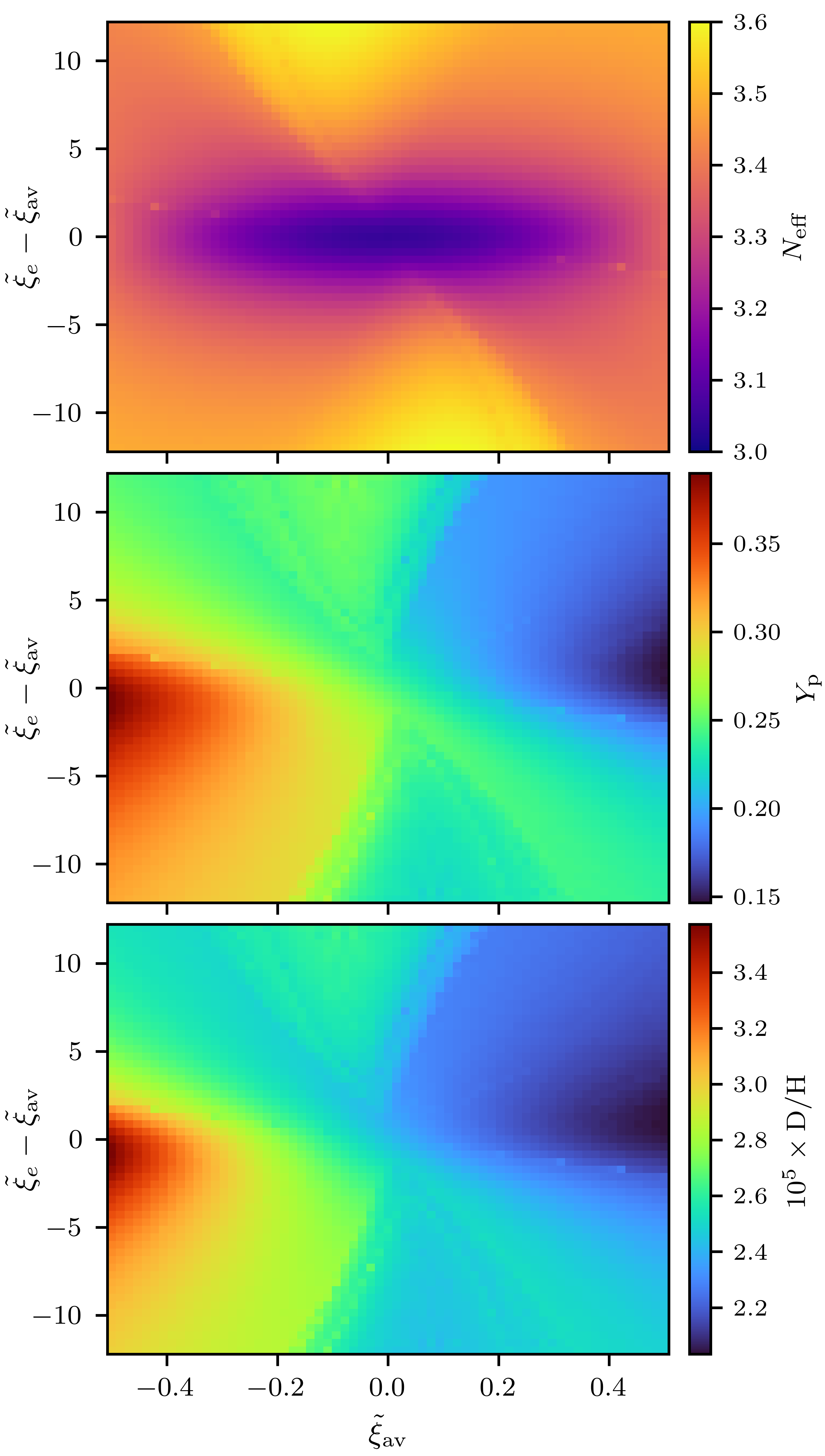}
    \caption{\label{fig:output_xie} Results of neutrino evolution and BBN calculation (for $\omega_b = \omega_b^\primat$) for equal $\nu_\mu$ and $\nu_\tau$ asymmetries. \emph{Top:} $\Neff$, invariant as expected through the transformation $(\xit_\av,\xit_e) \leftrightarrow (-\xit_\av,-\xit_e)$, corresponding to a central symmetry. \emph{Middle:} primordial helium-4 abundance $\Yp$. \emph{Bottom:} primordial deuterium abundance $\DH$. These abundances share similar patterns and are mostly determined by the final electron neutrino chemical potential; see Fig.~\ref{fig:etae_xie}.}
\end{figure}

\subsection{Equal \texorpdfstring{$\nu_\mu$}{νμ} and \texorpdfstring{$\nu_\tau$}{ντ} asymmetries}
\label{subsec:xie}

Since BBN is most sensitive to an electronic neutrino chemical potential (through the change in neutron-to-proton interconversion rates), it is justified to explore a range of values of $\xit_e$ different from the average by varying $\delta \xit_e$. In addition, we will change the average value of the asymmetries and will therefore show our results in the plane $(\xit_\av,\delta \xit_e)$, with a fixed value $\widetilde{\Delta} = 0$, which represents a minimal extension of the assumptions of Sec.~\ref{subsec:equal_asymmetries}. Such a value can be justified by the fact that synchronous flavor oscillations occur for the $\mu-\tau$ flavors earlier than $e-\mu$ and $e-\tau$ conversions (see, e.g., Fig.~\ref{fig:example_evolution}). If such oscillations brought initially different $\xi_\mu$ and $\xi_\tau$ to their average value, focusing on scenarios with $\Deltat = 0$ would be justified. However, this assumption is not true in general~\cite{Barenboim:2016shh},\footnote{Note that, if $\theta_{13} = 0$ and $\theta_{23} = \pi/4$ (maximal mixing), the $\mu$ and $\tau$ flavor distributions are indistinguishable at low temperatures.} as will be seen when we explore the effect of $\Deltat \neq 0$ in Sec.~\ref{subsec:delta}.

The results of our neutrino + BBN calculation on a regular grid of asymmetries, with $\xit_\av \in [-0.5,0.5]$ and $\delta \xit_e \in [-12,12]$, is shown on Fig.~\ref{fig:output_xie}. Generally, $\Neff$ increases with larger initial asymmetries, although the specific patterns are hard to predict. It is worth noting that a central symmetry, corresponding to the complete symmetry of neutrino evolution under the exchange $\nu_\alpha \leftrightarrow \bar{\nu}_\alpha$ for all three flavors, is satisfied. 

\begin{figure}[!ht]
    \centering
    \includegraphics{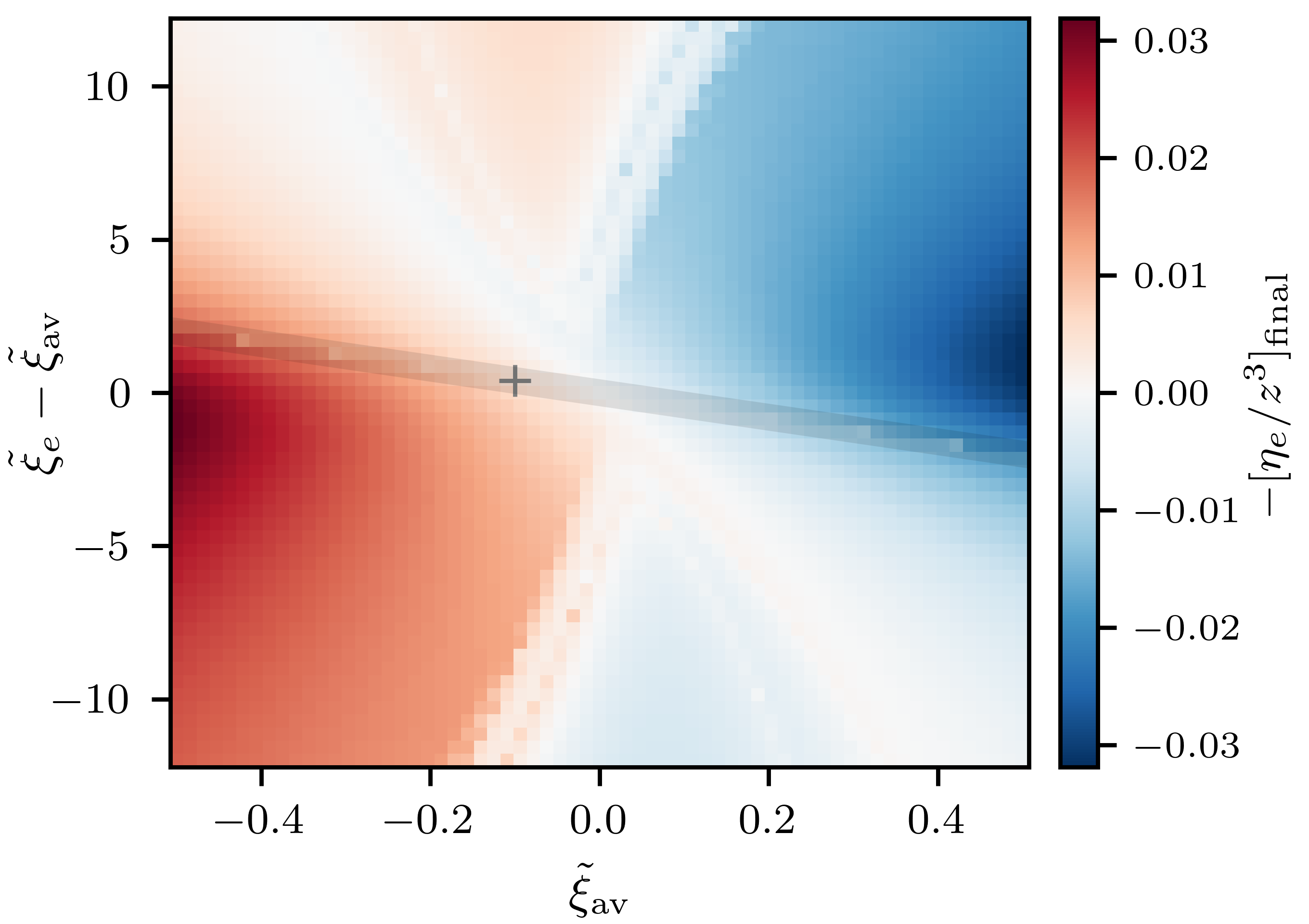}
    \caption{\label{fig:etae_xie} Final electron neutrino asymmetry, showing the correlation with the primordial abundances in Fig.~\ref{fig:output_xie}, as expected from Eq.~\eqref{eq:YpDHxi_old} or Eq.~\eqref{eq:YpDHxi}. The gray band corresponds to potential “equal but opposite asymmetries” situations, discussed in Sec.~\ref{subsec:EBO}. In particular, Fig.~\ref{fig:output_EBO} shows results on a zoomed-in grid around the point identified with a gray cross ($\xit_\av = -0.1$, $\delta \xit_e = 0.4$).}
\end{figure}

\begin{figure*}[!ht]
    \centering
    \includegraphics{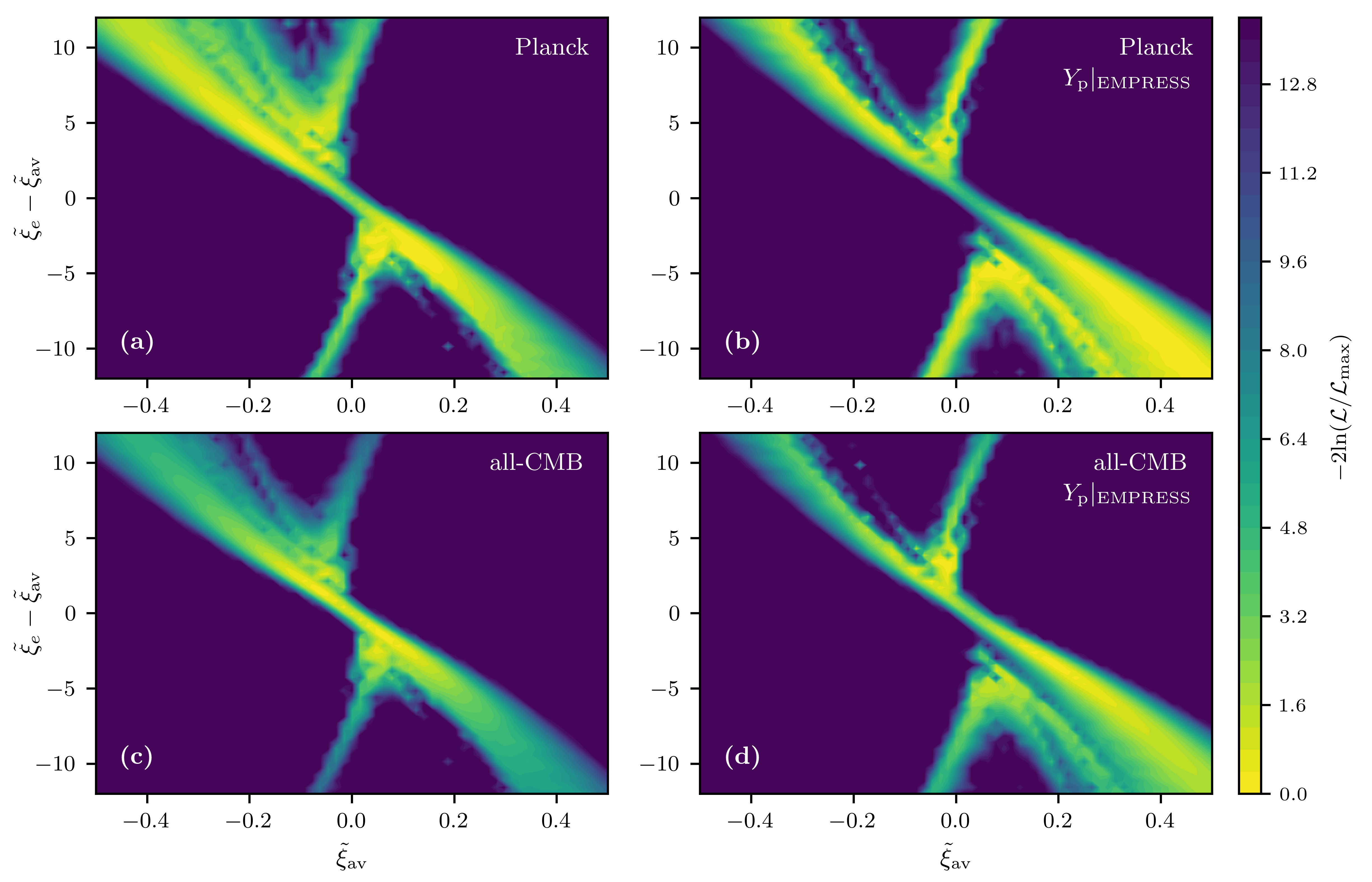}
    \caption{\label{fig:likelihood_all_xie} Total likelihood for initial asymmetries in the plane $(\xit_\av,\delta \xit_e)$, using the Aver et al.~value (left panels) or the EMPRESS value (right panels) of the helium-4 abundance. The CMB likelihood uses Planck + BAO data (top panels) or all current CMB experiments + BAO (bottom panels).}
\end{figure*}

Such a symmetry does not exist for the primordial abundances, since they are mainly modified by asymmetries through the change of neutron-to-proton ratio due to the electron neutrino asymmetry during BBN, $\xi_e^\mathrm{BBN}$. We have shown in Sec.~\ref{subsec:BBN} that a proxy for this quantity (which is not properly defined for the true out-of-equilibrium evolution of neutrinos) is the final value of $(\eta_e/z^3)$. It is represented on Fig.~\ref{fig:etae_xie} and shows the very same patterns as the abundance plots on Fig.~\ref{fig:output_xie}, confirming our physical understanding of the processes at play. We emphasize to the reader that assuming full flavor equilibration would mean that $\eta_e\rvert_\text{final} = \eta_\av$, thus Fig.~\ref{fig:etae_xie} would be invariant along the $y-$axis (see bottom panel of Fig.~\ref{fig:zmix_ximix}). This is not at all the case, with complicated patterns of variation of $\eta_e$ (and the other asymmetries) which result in the patterns observed on Fig.~\ref{fig:output_xie}.

\paragraph*{Likelihoods —} The result of the statistical analysis introduced in Sec.~\ref{subsec:likelihood} is shown on Fig.~\ref{fig:likelihood_all_xie}. The results of Sec.~\ref{subsec:equal_asymmetries} (e.g., Fig.~\ref{fig:likelihood_equalxi}) correspond to the line $\xit_e - \xit_\av = 0$ of the four panels. The case of equal asymmetries is \emph{not} representative of the general allowed space of $\xi_\alpha$. The shape of the likelihood is highly non-Gaussian (preventing the drawing of meaningful \SI{68}{\percent}, \SI{95}{\percent} contours), with high asymmetry points — i.e., $\xi_e = \mathcal{O}(10)$ — not excluded by our analysis. Using all CMB data compared to only Planck's data reduces the likelihood of these large $\xi$ points, but asymmetries of order unity are still allowed. As in the restricted case of equal asymmetries, using the EMPRESS measurement of the helium-4 abundance pushes the likelihood away from the zero asymmetry point $(0,0)$.

Comparison of the likelihood (Fig.~\ref{fig:likelihood_all_xie}) and physical input (Fig.~\ref{fig:output_xie}) patterns shows that the total likelihood is mostly determined by the BBN measurements, which is expected given the much larger relative uncertainties on $\Neff,\Yp$ from CMB [Eqs.~\eqref{eq:means_Planck} and \eqref{eq:means_allCMB}] compared to the percent level spectroscopic measurements of $\Yp,\DH$ [Eqs.~\eqref{eq:Yp_mes}--\eqref{eq:Yp_empress}].

\subsection{Varying \texorpdfstring{$\xi_\mu - \xi_\tau$}{ξμ - ξτ}}
\label{subsec:delta}

We now explore the role of initial differences between the $\mu$ and $\tau$ flavor asymmetries. We assume that initially $\xi_e = \xi_\av$. The example of Fig.~\ref{fig:example_evolution}, which corresponds to such parameters, highlights the need for the dynamical evolution of asymmetries, as the final values of $\eta_\alpha$ cannot be straightforwardly obtained from the initial values.

\begin{figure}[!ht]
    \centering
    \includegraphics{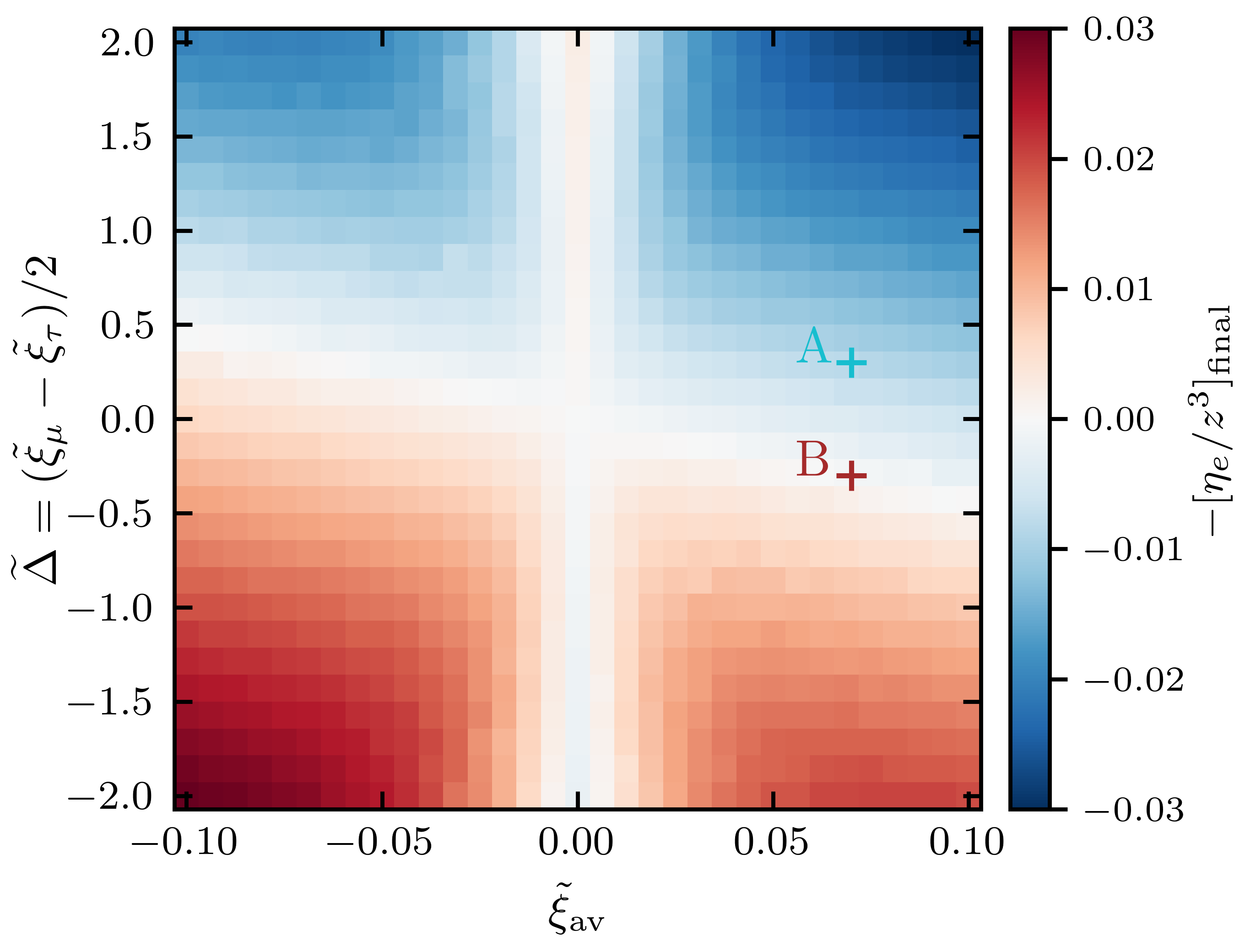}
    \caption{\label{fig:etae_delta} Final electron neutrino asymmetry in the plane $(\xit_\av,\Deltat)$. The patterns show that flavor equilibration is not a common result at all, as this plot would then be invariant vertically (because $\eta_e\rvert_\mathrm{final}$ would be equal to $\eta_\av$). Two points, which only differ by opposite values of $\widetilde{\Delta}$, are highlighted by crosses and the associated time evolution of asymmetries is shown on Fig.~\ref{fig:compare_output_delta}.}
\end{figure}

\begin{figure}[!ht]
    \centering
    \includegraphics{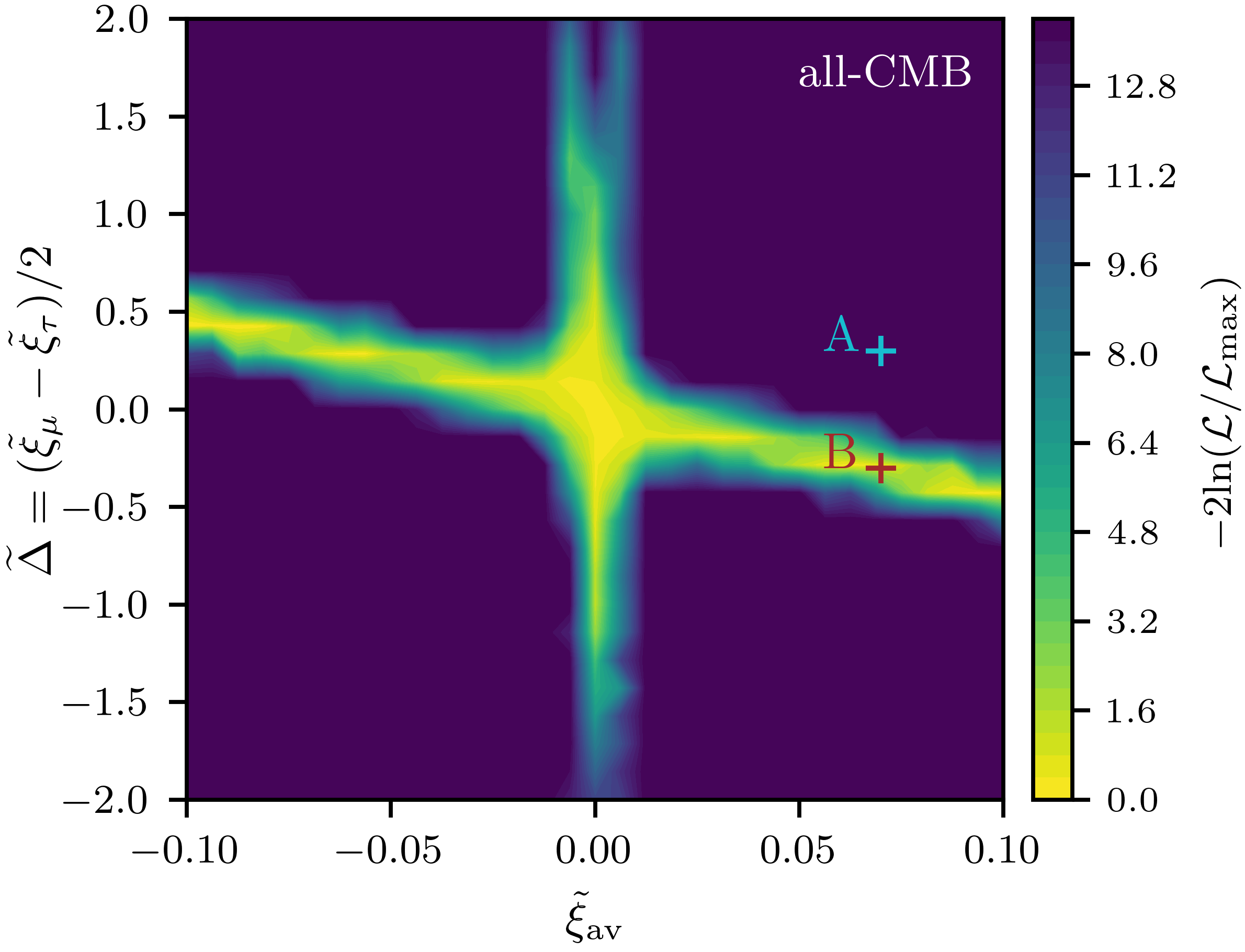}
    \caption{\label{fig:likelihood_delta} Total likelihood in the plane $(\xit_\av,\widetilde{\Delta})$, using all-CMB likelihood and the Aver et al. value of the helium abundance \eqref{eq:Yp_mes}. Large regions are allowed; they correspond to final values of the electron flavor asymmetry close to zero (see Fig.~\ref{fig:etae_delta}), resulting in primordial abundances only slightly modified compared to the zero asymmetry case.}
\end{figure}

We run our calculations for initial asymmetries $\{\delta \xit_e = 0, \Deltat \neq 0\}$, equally spaced on the plane $(\xit_\av,\Deltat)$, with $\xit_\av \in [-0.1,0.1]$ and $\Deltat \in [-2,2]$. The proxy for $\xi_e^\mathrm{BBN}$ on this grid of asymmetries, similarly to Fig.~\ref{fig:etae_xie}, is shown on Fig.~\ref{fig:etae_delta}, and the likelihood is shown on Fig.~\ref{fig:likelihood_delta}. Similarly to the results of Fig.~\ref{fig:likelihood_all_xie}, we see that current experimental data do not allow one to set good constraints on the asymmetries, as the high likelihood regions extend beyond the explored grid. We expect that high-$\xit_\av$ points would be rejected with a better precision on the experimental determination of $\Neff$, as will be shown in Sec.~\ref{sec:forecast}. This once again highlights the limitations of assuming flavor equilibration: the patterns on Fig.~\ref{fig:etae_delta} show explicitly that the final asymmetries are not equal in general, as there would otherwise be a symmetry via $\Deltat \to - \Deltat$. 

 We can identify two “high likelihood” regions: the vertical band $\xit_\av \simeq 0$, and an oblique band (roughly given by $\Deltat \simeq -0.5 \, \xit_\av$). They correspond to regions where the final electron flavor asymmetry is close to zero (see Fig.~\ref{fig:etae_delta} and the previous discussion in Sec.~\ref{subsec:xie}). For the vertical band, this corresponds to the fact that initially $\xi_e = 0$, and the value remains small after flavor transformation. Regarding the oblique band, these results underline once again why a precise neutrino evolution code is needed to draw meaningful conclusions. Notably, the final electron flavor asymmetry does not follow the patterns of Fig.~\ref{fig:zmix_ximix} (bottom panel), showing that flavor equilibration is not satisfied. 
 
\begin{figure}[!ht]
    \centering
    \includegraphics{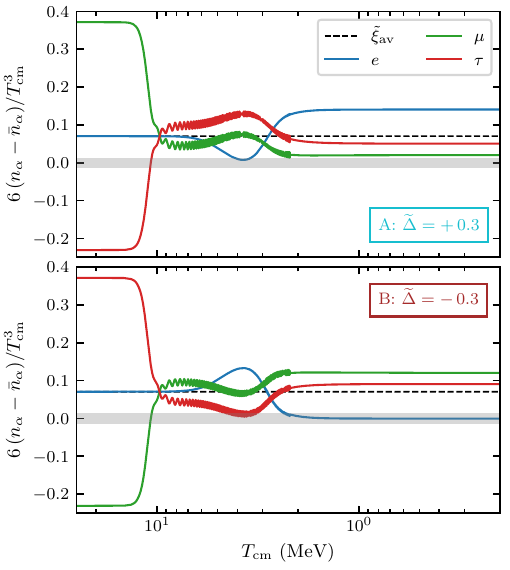}
    \caption{\label{fig:compare_output_delta} Evolution of neutrino asymmetries for the conditions of points A and B identified on Fig.~\ref{fig:likelihood_delta}. Initially $\xit_e = \xit_\av = 0.07$, the only difference being the sign of $\widetilde{\Delta}$. \emph{Top:} Point A, $\widetilde{\Delta} = + 0.3$. The electron flavor asymmetry during BBN is large, leading to primordial abundances incompatible with measurements. \emph{Bottom:}  Point B, $\widetilde{\Delta} = - 0.3$. The final electron flavor asymmetry is close to zero, explaining why Point B is in the “allowed” region.}
\end{figure}

 To further illustrate this feature, we identify on Fig.~\ref{fig:likelihood_delta} two points (“Point A” and “Point B,” with $\xit_\av^{(\text{A})} = \xit_\av^{(\text{B})} = 0.07$, $\Deltat^{(\text{A})} = + 0.3$ and $\Deltat^{(\text{B})}=-0.3$), which only differ by the sign of $\Deltat$, that is, by the exchange $(\nu_\mu, \bar{\nu}_\mu) \leftrightarrow (\nu_\tau,\bar{\nu}_\tau)$. The time evolution of asymmetries for these two points is shown on Fig.~\ref{fig:compare_output_delta}. The high likelihood for Point B compared to Point A is due to the final electron flavor asymmetry, which is very small for Point B (Fig.~\ref{fig:compare_output_delta}, bottom panel) but not for Point A (Fig.~\ref{fig:compare_output_delta}, top panel). Since, for now, quantities are mostly constrained through the primordial abundances and not CMB data, i.e., the value of $\Neff$ is poorly constrained, the set of parameters $\{\xit_\av, \delta \xit_e, \Deltat \}^{(\text{B})}$ is allowed. However, a final value $\eta_e \simeq 0$ means, by conservation of the average $\hat{\eta}$, that the final asymmetries $\eta_\mu$ and $\eta_\tau$ can be large if $\xit_\av$ is large. This leads to a higher value of $\Neff$, which should be excluded by future, more precise, CMB experiments (see Sec.~\ref{sec:forecast}).

\subsection{Equal but opposite asymmetries}
\label{subsec:EBO}

The vertical line of high likelihood ($\xit_\av \simeq 0$) on Fig.~\ref{fig:likelihood_delta}, which corresponds to initial parameters $\{\xi_e = 0, \xi_\mu = - \xi_\tau\}$, is a case of “equal but opposite” (EBO) asymmetries~\cite{Dolgov_NuPhB2002,Froustey2021}. Such configurations lead to a delayed onset of (quasi)synchronous oscillations.\footnote{The regime of \emph{quasi-}synchronous oscillations occurs when $\H_0 + \H_\mathrm{lep}$ is not negligible anymore compared to $\Hself$. Due to the different signs in the Hamiltonians of Eqs.~\eqref{eq:QKE_rho} and \eqref{eq:QKE_rhobar}, the perfect synchronicity of neutrino and antineutrino collective oscillations is broken, which results in an accelerated frequency for the evolution of the asymmetry~\cite{Froustey2021}.} For instance, for a two-flavor system, the cancellation of the sum of asymmetries prevents synchronous oscillations from happening. Only quasi-synchronous oscillations can take place, but as they are a higher order effect in $\lVert\H_0 + \H_\mathrm{lep}\rVert/\lVert \Hself \rVert$, oscillations are delayed to lower temperatures. Such configurations make dynamical calculations more challenging, as we must be certain to follow the evolution until oscillations are damped enough to get to an \ATAOH regime. This is not a problem for the calculations of Sec.~\ref{subsec:delta}: as oscillations in the $\{\nu_\mu-\nu_\tau\}$ subspace usually start around $\Tcm \sim \SI{10}{\mega \electronvolt}$, the delay due to the EBO configurations is still readily manageable.

\begin{figure}[!ht]
    \centering
    \includegraphics{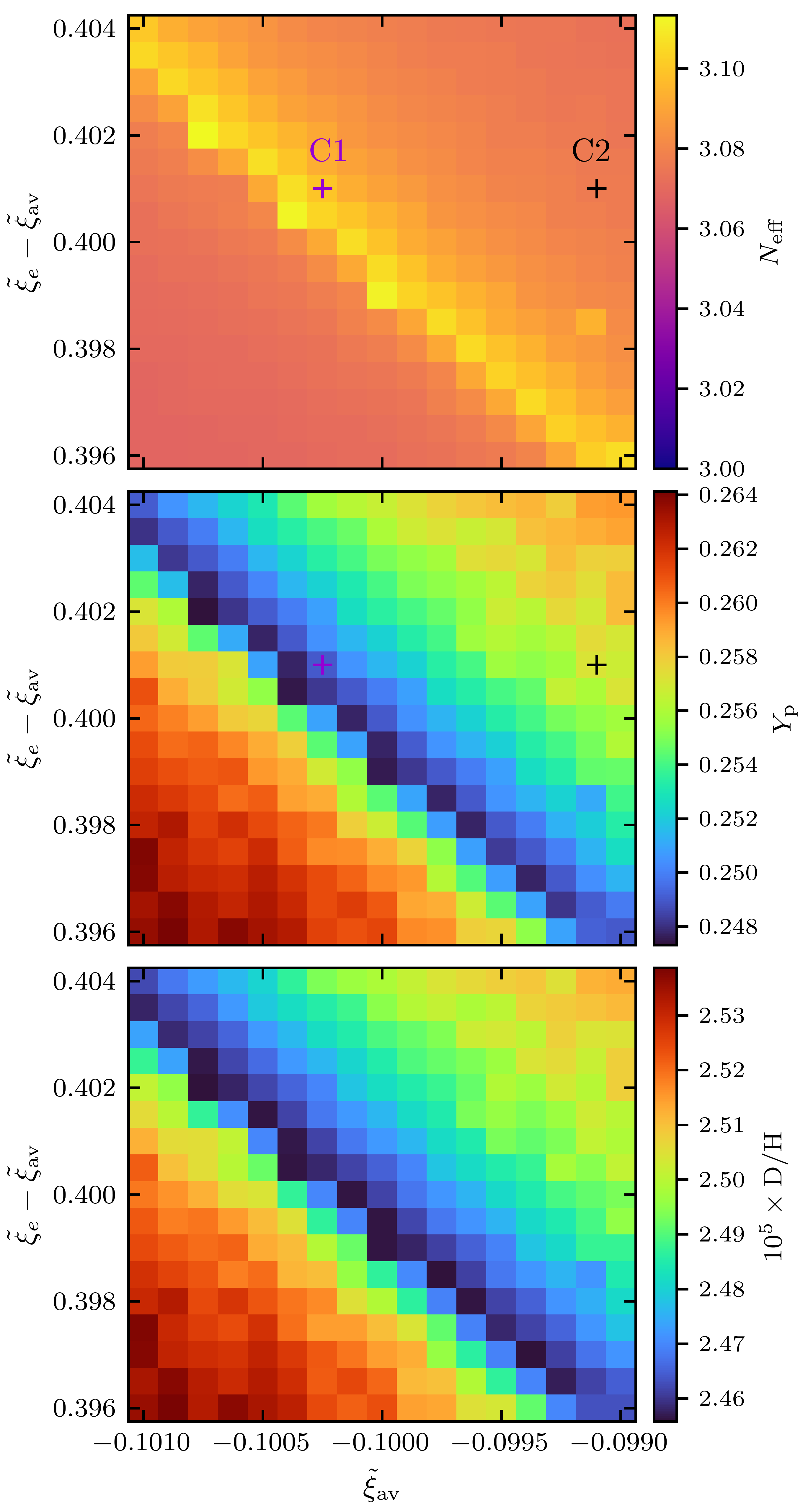}
    \caption{\label{fig:output_EBO} Results of neutrino evolution and BBN calculation for equal $\nu_\mu$ and $\nu_\tau$ asymmetries, around a point of “equal but opposite” asymmetries (see gray cross on Fig.~\ref{fig:etae_xie}). For these calculations, the default transition temperature between \ATAOJH and \ATAOH schemes is set at $\bar{T}_\trans = \SI{1.6}{\mega \electronvolt}$ (see Sec.~\ref{subsec:Nu_transport}). Plotting conventions are the same as Fig.~\ref{fig:output_xie}.}
\end{figure}

However, when initially $\Deltat = 0$, conversions can only take place with the $\nu_e$ flavor, which occurs for smaller temperatures. In the plane $(\xit_\av, \delta \xit_e)$, with $\Deltat=0$ (see Sec.~\ref{subsec:xie}), this situation arises for initial asymmetries satisfying $\xi_\mu = \xi_\tau = - \xi_e$. Considering $\xi \simeq \xit$ for small asymmetries and using the relations~\eqref{eq:parametrization}, we see that these “pathological” points are on the line
\begin{equation}
\label{eq:line_EBO}
    \delta \xit_e = - 4 \xit_\av \, .
\end{equation}
This line is shown as a gray band on Fig.~\ref{fig:etae_xie}. One can see, on Figs.~\ref{fig:output_xie} and \ref{fig:etae_xie}, that a few grid points along this line show results seemingly with artifacts. We investigate this feature by exploring a small region of parameters around the point $(\xit_\av,\delta \xit_e) = (-0.1,0.4)$ — this point is shown with a gray cross on Fig.~\ref{fig:etae_xie}. The results of our neutrino + BBN calculation are shown on Fig.~\ref{fig:output_EBO}.

\begin{figure}[!t]
    \centering
    \includegraphics{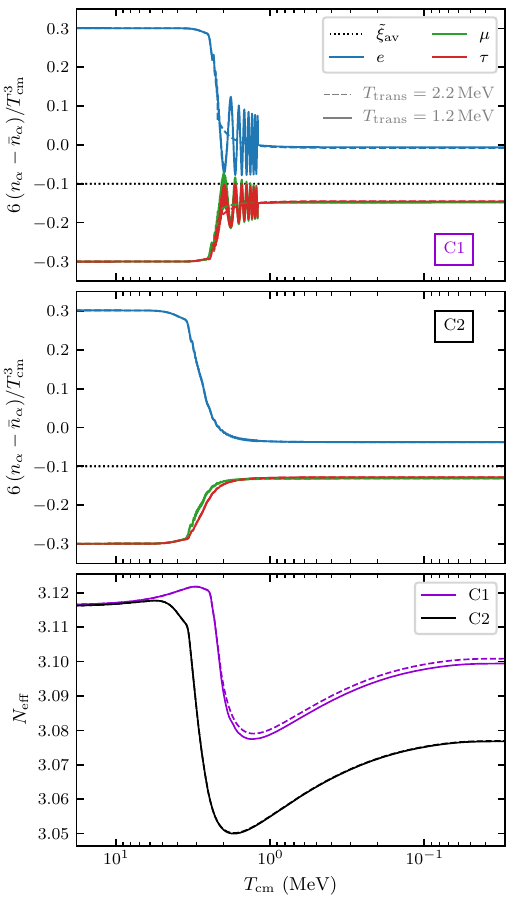}
    \caption{\label{fig:output_C1C2} Neutrino evolution for the conditions of points C1 and C2 identified on Fig.~\ref{fig:output_EBO}. We consider two transition temperatures between the \ATAOJH and \ATAOH schemes in \nevo: $T_\trans = \SI{2.2}{\mega \electronvolt}$ (dashed lines) and $T_\trans = \SI{1.2}{\mega \electronvolt}$ (solid lines). \emph{Top:} evolution of asymmetries for point C1. \emph{Middle:} evolution of asymmetries for point C2. \emph{Bottom:} $\Neff$, as given by Eq.~\eqref{eq:Neff_gen}, for each calculation.}
\end{figure}

One can observe, along the line given by Eq.~\eqref{eq:line_EBO}, apparently peculiar results, which do not seem to follow the background trend, whether this is for $\Neff$, $\Yp$, or $\DH$. Regarding primordial abundances, the reasons come from the coincidence between this line and a range of parameters that make the final electron flavor asymmetry very small. For $\Neff$, this is actually a feature of EBO configurations, for which the onset of collective oscillations is delayed to lower temperatures compared to neighboring configurations. We illustrate this property on Fig.~\ref{fig:output_C1C2}, where we compare the time evolution of asymmetries for two points (“C1” and “C2”) highlighted on Fig.~\ref{fig:output_EBO}, top panel. Point C1 ($\xit_\av = -0.10025, \delta \xit_e = 0.401$) is exactly on the line~\eqref{eq:line_EBO}, while point C2 ($\xit_\av = - 0.0991, \delta \xit_e = 0.401$) is close to it. The dynamical behavior of the asymmetries is very different in the two cases. First, the onset of oscillations, which corresponds to the Mikheyev–Smirnov–Wolfenstein (MSW) transition from matter-dominated to vacuum-dominated in the Hamiltonian “felt” by the asymmetry~\cite{Froustey2021}, is delayed from C2 ($\Tcm \sim \SI{3}{\mega \electronvolt}$) to C1 ($\Tcm \sim \SI{2}{\mega \electronvolt}$). This delay is responsible for the nonadiabaticity of the transition experienced by the asymmetry, which results in large oscillations for C1. This shows the importance of following the full Hamiltonian [\ATAOJH scheme] to low enough temperatures: we represent on Fig.~\ref{fig:output_C1C2} in solid lines (resp.~dashed lines) the evolution using $T_\trans = \SI{1.2}{\mega \electronvolt}$ (resp.~$T_\trans = \SI{2.2}{\mega \electronvolt}$). For C1, the higher $T_\trans$ case does not represent accurately the transition, although the average values are well described, hence a limited impact on the final observables. However, for other points where the onset of oscillations would be pushed to even lower temperatures, a premature shift from \ATAOJH to \ATAOH may cause some differences. This is notably evidenced when we focus on the energy density budget between (anti)neutrinos and photons. 

To depict the transfers of energy due to flavor mixing and later reheating by electron-positron annihilations, we define the effective number of neutrinos at any temperature as
\begin{equation}
\label{eq:Neff_gen}
    \Neff(\Tcm) \equiv \frac{\sum_{\alpha}{(\rho_\alpha + \bar{\rho}_\alpha})}{\frac{7 \pi^2}{120}\Tcm^4} \times \left(\frac{z^{(0)}}{z}\right)^4 \, ,
\end{equation}
where $z^{(0)}$ is the temperature that photons would have in the instantaneous decoupling approximation, without QED corrections, such that $z^{(0)}(\Tcm \gg m_e) = 1$ and $z^{(0)}(\Tcm \ll m_e) = (11/4)^{1/3}$. It can be obtained, for instance, via entropy conservation. The value of \eqref{eq:Neff_gen} for $\Tcm \ll m_e$ corresponds to the quantity $\Neff$ shown on the top panels of Figs.~\ref{fig:output_equalxi}, \ref{fig:output_xie} and \ref{fig:output_EBO}. 

This generalized $\Neff$ is shown on the bottom panel of Fig.~\ref{fig:output_C1C2}. Consistently with Fig.~\ref{fig:output_EBO}, $\Neff\rvert_\mathrm{C1} > \Neff\rvert_\mathrm{C2}$. We interpret this as a consequence of the EBO configuration: when asymmetries mix, the extra energy density contained in (anti)neutrinos is redistributed as a common temperature for (anti)neutrinos, electrons, positrons and photons. However, if this partial equilibration occurs later, the thermal contact between (anti)neutrinos and the electromagnetic plasma is gradually broken and photons get, in comparison, a lower share of the redistributed energy. This leads to a higher value for $\Neff$. Note, in addition, that the final increase due to $e^- e^+$ annihilations, for $\Tcm \leq \SI{1}{\mega \electronvolt}$, is almost identical for C1 and C2. For the EBO configuration, using a transition temperature that is too high (here, \SI{2.2}{\mega \electronvolt}, see dashed lines) leads to a small error in $\Neff$, which could be even larger in other EBO configurations if the onset of oscillations is delayed to even further temperatures.

To conclude, the patterns on the EBO line seen on Figs.~\ref{fig:output_xie} and~\ref{fig:etae_xie} are not mere artifacts and are resolved with our code. However, due to the computationally challenging nature of these configurations, the accuracy of our results on such points is lower compared to other configurations. We see, however, that this line is not favored by our statistical analysis (see Fig.~\ref{fig:likelihood_all_xie}), which makes this accuracy loss only marginally relevant.

\section{Forecasts}
\label{sec:forecast}

We have seen that, given the current CMB experiments, primordial neutrino asymmetries are very loosely constrained when one does not restrict to equal asymmetries for all three flavors. However, the improved capabilities of future CMB experiments will provide better constraints, which we estimate in this section.

\subsection{Upcoming CMB experiments}

We focus on two future CMB experiments: the Simons Observatory~\cite{SimonsObservatory:2018koc,SimonsObservatory:2019qwx}, an array of three 0.4-metre small-aperture telescopes and one 6-metre large aperture telescope in the Atacama desert, in the final stages of construction; and a next-generation project like CMB-Stage IV~\cite{CMB-S4:2016ple,Abazajian:2019eic}. They notably aim at constraining $\Neff$ at the percent level.

We provide forecasts for the capabilities of these telescopes following the strategy of~\cite{Sabti:2019mhn,Escudero:2022okz}. We use the baseline covariance matrices of Simons Observatory (SO) and CMB-Stage IV (CMB-S4):
\begin{align}
\label{eq:cov_SO}
\Sigma_\mathrm{SO} &= \begin{pmatrix}
    \num{5.33e-9} & \num{5.78e-7} & \num{1.59e-7} \\
    \num{5.78e-7} & \num{0.0121} & - 0.000624 \\
    \num{1.59e-7} & -0.000624 & \num{4.36e-5}
\end{pmatrix} \, , \\
\label{eq:cov_CMBS4}
\Sigma_\text{CMB-S4} &= \begin{pmatrix}
    \num{2.21e-9} & \num{9.52e-7} & \num{4.45e-8} \\
    \num{9.52e-7} & 0.00656 & - 0.000293 \\
    \num{4.45e-8} & -0.000293 & \num{1.85e-5}
\end{pmatrix} \, .
\end{align}

\begin{figure}[!ht]
    \centering   \includegraphics{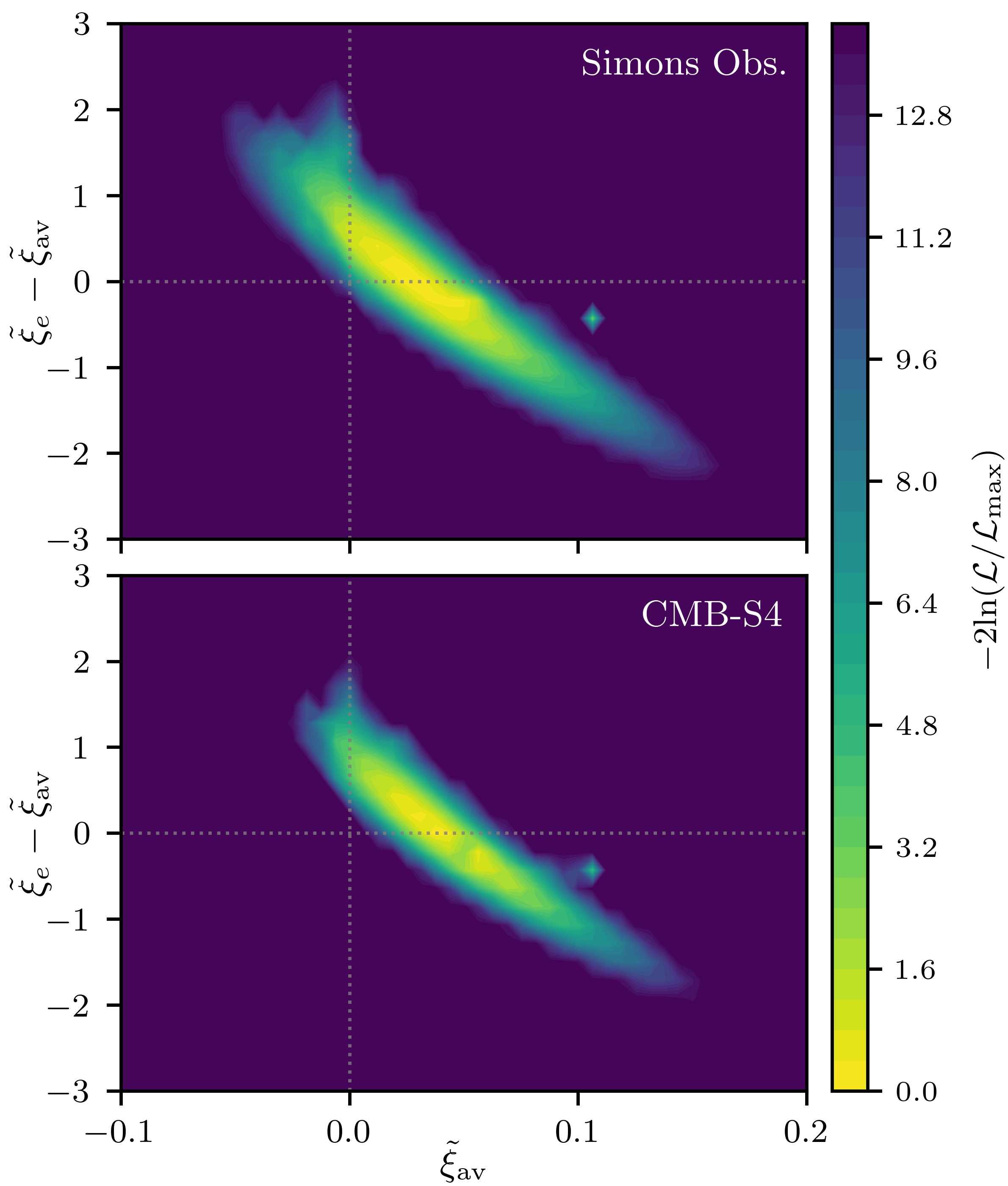}  \caption{\label{fig:likelihood_forecast_xie} Forecast of the total likelihood for initial asymmetries in the plane $(\xit_\av,\delta \xit_e)$, using the Aver et al. value of the helium-4 abundance~\eqref{eq:Yp_mes}. \emph{Top:} results using the SO covariance matrix for the CMB likelihood $\mathcal{L}_\mathrm{CMB}$. \emph{Bottom:} results using the CMB-S4 covariance matrix for $\mathcal{L}_\mathrm{CMB}$. The central values in $\mathcal{L}_\mathrm{CMB}$ are set to the values of Eq.~\eqref{eq:means_allCMB}.}
\end{figure}

\subsection{Improved constraints}

We show on Fig.~\ref{fig:likelihood_forecast_xie} the likelihood obtained in the plane $(\xit_\av,\delta \xit_e)$, using the covariance matrices \eqref{eq:cov_SO} and \eqref{eq:cov_CMBS4} for the CMB likelihood and setting the central values of $\mathcal{L}_\mathrm{CMB}$ to the same values as the ones obtained when combining all existing CMB experiments, see Eq.~\eqref{eq:means_allCMB}. Note that, compared to Fig.~\ref{fig:likelihood_all_xie}, the span of asymmetries is reduced as these upcoming CMB experiments provide much tighter constraints on the asymmetries. Some “peculiar” points, which correspond to EBO configurations (see Sec.~\ref{subsec:EBO}), can be noticed.

The zero asymmetry configuration $\{\xi_e=\xi_\mu=\xi_\tau=0\}$ would appear to be disfavored. However, this result depends largely on the central values adopted for the CMB likelihood. In Fig.~\ref{fig:likelihood_forecast_xie}, we have taken the same means as the ones currently obtained in CMB experiments [Eq.~\eqref{eq:means_allCMB}]. With these new covariance matrices, the standard deviation on each parameter is typically reduced by a factor of 10, which makes the future determination of $\Yp$ competitive with current spectroscopic measurements. However, the central value~\eqref{eq:mean_Yp_allCMB} is lower than \eqref{eq:Yp_mes} and pushes for nonzero asymmetries, similarly to the lower spectroscopic value from EMPRESS~\eqref{eq:Yp_empress}. 

\begin{figure}[!ht]
    \centering    \includegraphics{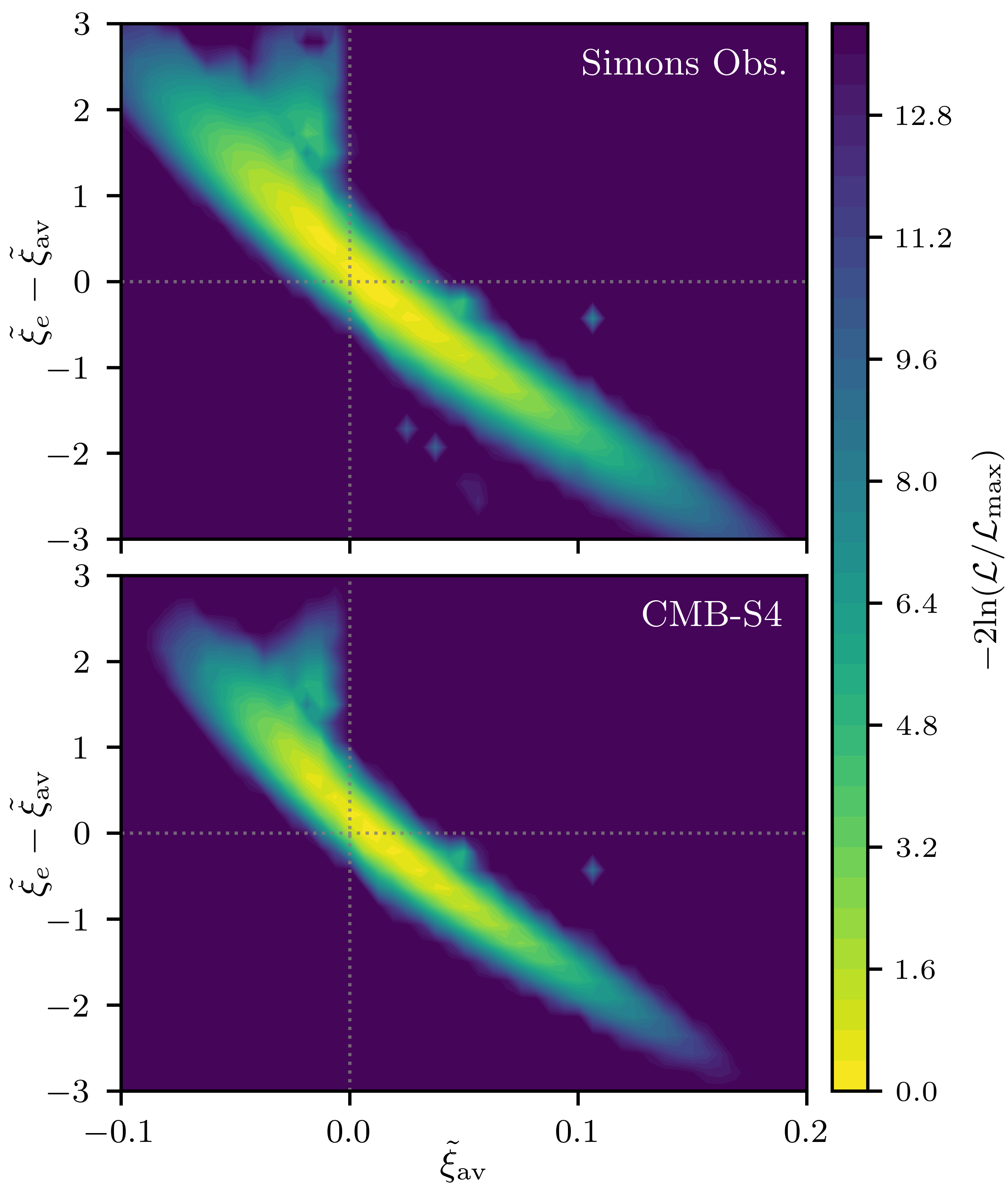}
    \caption{\label{fig:likelihood_forecast_xie_Ypmean} Same as Fig.~\ref{fig:likelihood_forecast_xie} but setting the central value of $\Yp$ in $\mathcal{L}_\mathrm{CMB}$ to \eqref{eq:Yp_mes}. Vanishing asymmetries are not disfavored anymore with this choice.}
\end{figure}

As a comparison, we show on Fig.~\ref{fig:likelihood_forecast_xie_Ypmean} the same results as Fig.~\ref{fig:likelihood_forecast_xie}, but setting the central value of $\Yp$ in the CMB likelihood to the Aver et al. value~\eqref{eq:Yp_mes}. With this new choice, the zero asymmetry configuration is not disfavored anymore. The shape of the likelihood is different between Figs.~\ref{fig:likelihood_forecast_xie} and \ref{fig:likelihood_forecast_xie_Ypmean}, but they are still approximately Gaussian and closed: future CMB experiments will provide significant constraints on primordial neutrino asymmetries.

\section{Conclusion}
\label{sec:conclusion}

We have revisited the constraints on cosmological lepton asymmetries in light of the recent progress on the determination of mixing parameters and on the numerical description of neutrino evolution in the early Universe, using a three-flavor multi-momentum code with the full neutrino collision term and the BBN code \primat.

Constraints on asymmetries at the BBN epoch, commonly established in the literature, are not representative of the constraints on \emph{primordial} asymmetries. In particular, since total flavor equilibration is not verified in general, it is not straightforwardly possible to connect initial and final asymmetries, which shows the need for an accurate dynamical evolution of (anti)neutrinos through the MeV age. Still, our results are consistent with the studies that focus on BBN-epoch asymmetries; see the constraints~\eqref{eq:Equal_xi} obtained by assuming equal primordial asymmetries — in which case they do not evolve and they match the values at the BBN epoch. For future use, we also obtained an updated semianalytical expression of the dependence of primordial abundances on the electron neutrino chemical potential at BBN [Eq.~\eqref{eq:YpDHxi}].

It is well known that, because of flavor oscillations, a large space of primordial asymmetries can be averaged to small asymmetries consistent with BBN, but our study quantifies this with a reliable code that does not approximate the collision term, allowing for dedicated future studies of particular lepton asymmetry generation models. Our main finding is that current CMB experiments do not allow one to constrain these primordial asymmetries, with generically unbound domains of the parameter space favored by our statistical analysis. This is due to the fact that the limits on lepton asymmetries are dominated by the percent-level measurements of primordial abundances, which are compatible with large domains of the parameter space that lead to a vanishing final electron flavor asymmetry. In other words, the constraints on $\Neff$ are not strong enough to discard large initial asymmetries that lead to a very small $\xi_e^\mathrm{BBN}$.

We highlighted a potential limit of our analysis, which corresponds to the cases of “equal but opposite asymmetries.” These peculiar configurations are associated with a delay of the onset of collective oscillations, which are harder to capture accurately. Although we have shown that these specific situations do not challenge our conclusions, a dedicated study of these configurations, should they arise in a specific model of lepton asymmetry generation, can readily be carried out.

Finally, we assessed the potential of future CMB experiments like the Simons Observatory, which aim at determining $\Neff$ (and $\Yp$) with a precision better by an order of magnitude compared to current CMB experiments. They should be able to “close the contours” and provide significant constraints on the primordial asymmetries, which will greatly inform models of baryogenesis and leptogenesis.

\section*{Acknowledgments}

We thank Silvia Galli, Karim Benabed, and Ali Rida Khalife for their help with statistics and the SPT-3G likelihood. We thank Joel Meyers for useful discussions on the statistical analysis of future CMB experiments and for providing the baseline covariance matrix of the Simons Observatory. We are grateful to Cynthia Trendafilova for providing the Fisher matrix for CMB-S4 preliminary baseline design configuration. This work has made use of the Infinity cluster hosted by the Institut d'Astrophysique de Paris. J.F. thanks the Institut d'Astrophysique de Paris for its hospitality during various stages of this work. J.F. is supported by the Network for Neutrinos, Nuclear Astrophysics and Symmetries (N3AS), through the National Science Foundation Physics Frontier Center Award No.~PHY-2020275.

\appendix

\section{Statistical analysis}
\label{App:statistics}

The experimental data we are using to constrain neutrino properties are the CMB power spectrum ($C_\ell^\obs$) and the primordial abundances ($\Yp^\obs$ and $\DH^\obs$). We want to determine the probability of the theory parameters $\{\xi_\alpha,\omega_b\}$ given these measurements, that is:
\begin{multline}
\label{eq:likelihood_withom}
    \L(\xi_\alpha, \omega_b \rvert C_\ell^\obs,\Yp^\obs,\DH^\obs) \\ \sim P(C_\ell^\obs,\Yp^\obs,\DH^\obs \rvert \xi_\alpha, \omega_b) \, ,
\end{multline}
through Bayesian inversion. The final conditional probability can be separated between CMB and BBN measurements, since they are independent.

Let us start with BBN measurements. The conditional probability to measure $\{\Yp^\obs,\DH^\obs\}$ knowing the “true” values $\{\Yp,\DH\}$ is given by the product of likelihoods
\begin{multline}
    P(\Yp^\obs,\DH^\obs \rvert \Yp, \DH) = \mathcal{N}(\Yp^\obs ; \Yp, \sigma_{\Yp,\obs}) \\
    \times \mathcal{N}(\DH^\obs ; \DH, \sigma_{\DH,\obs}) \, ,
\end{multline}
with the Gaussian likelihood being, in general,
\begin{equation}
\mathcal{N}(A^\obs;A,\sigma_A) = \frac{1}{\sqrt{2 \pi} \sigma_A} e^{-(A^\obs - A)^2/2 \sigma_A^2} \, .
\end{equation}
We relate this quantity to the theory parameters $\{\xi_\alpha, \omega_b\}$ via, for instance for helium-4,
\begin{equation}
P(\Yp^\obs \rvert \xi_\alpha, \omega_b) = \int{\dd{\Yp} P(\Yp^\obs \rvert \Yp) \times P(\Yp \rvert \xi_\alpha, \omega_b)} \, .
\end{equation}
The last conditional probability is actually the output from our codes \nevo $+$ \primat. For a given set of theory parameters, the final result has a numerical uncertainty due mainly to the uncertainties on the nuclear rates and the neutron lifetime. This variation is estimated via a Monte-Carlo search in \primat, and we find $\sigma_{\Yp,\code}/\Yp \simeq 0.051 \%$ and $\sigma_{\DH,\code}/(\DH) \simeq 1.13 \%$. Therefore, the overall result for BBN measurements reads

\begin{widetext}
\begin{align}
    P(\Yp^\obs,\DH^\obs \rvert \xi_\alpha, \omega_b) &= \int{\dd{\Yp} \dd{(\DH)} \, P(\Yp^\obs,\DH^\obs \rvert \Yp, \DH) \times P(\Yp \rvert \xi_\alpha, \omega_b) P(\DH \rvert \xi_\alpha, \omega_b)} \\
    &= \int{\dd{\Yp} \dd{(\DH)} \, \mathcal{N}(\Yp^\obs ; \Yp, \sigma_{{\Yp},\obs}) \times \mathcal{N}(\DH^\obs ; \DH, \sigma_{{\DH},\obs})} \\
    &\qquad \qquad \qquad \qquad \times \mathcal{N}(\Yp ; {\Yp}^\code_{(\xi_\alpha,\omega_b)}, \sigma_{{\Yp},\code}) \times \mathcal{N}(\DH ; \DH^\code_{(\xi_\alpha,\omega_b)}, \sigma_{\DH,\code}) \nonumber \\
    &= \mathcal{N}\left(\Yp^\obs ; {\Yp}^\code_{(\xi_\alpha,\omega_b)}, \sigma_{\Yp}\right) \times \mathcal{N}\left(\DH^\obs ; {\DH}^\code_{(\xi_\alpha,\omega_b)}, \sigma_{\DH}\right)
\end{align}
where, by convolution of Gaussian distributions, the total standard deviation is $\sigma_{\Yp} = \sqrt{\sigma_{{\Yp},\obs}^2 + \sigma_{{\Yp},\code}^2}$ and likewise for deuterium.
\end{widetext}

\begin{figure}[!ht]
    \centering
    \includegraphics[width=\columnwidth]{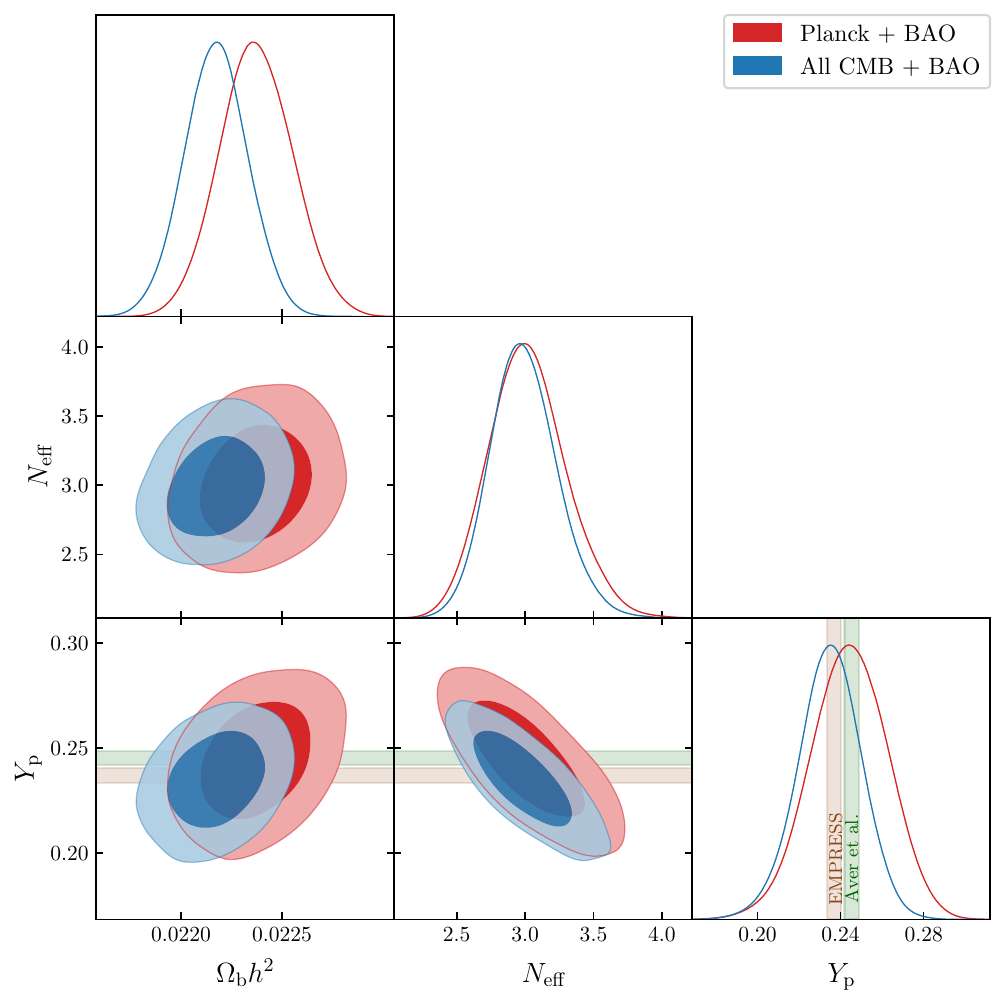}
    \caption{\label{fig:triangle_CMB} $68\%$ and $95\%$ contours of the posterior distribution in the $\Lambda$CDM model with additional free parameters $N_{\rm eff}$ and $Y_{\rm p}$. “All CMB + BAO” contours use data from BICEP/Keck~\cite{BICEP:2021xfz}, ACT~\cite{ACT:2020frw,ACT:2020gnv,Carron:2022eyg,ACT:2023dou,ACT:2023kun}, and SPT-3G~\cite{SPT-3G:2022hvq} in addition to Planck~\cite{Planck:2019nip} and BAO~\cite{eBOSS:2020yzd} likelihoods. On the bottom right panel, colored bands show the helium-4 abundance measurements \eqref{eq:Yp_mes} and \eqref{eq:Yp_empress}.
    }
\end{figure}

Concerning CMB measurements, the CMB likelihood is expressed in terms of $\omega_b$, $\Neff$, and $\Yp$. The uncertainties on $\Neff$ from \nevo being much smaller than the ones on primordial abundances from \primat, we neglect them and consider that there is a unique correspondence $P(\Neff \rvert \xi_\alpha, \omega_b) = \delta(\Neff - \Neffxi^\code)$. Note in particular that the numerical prediction for $\Neff$ does not depend on the baryon density $\omega_b$, since the baryon-to-photon ratio is too small to have a measurable effect on lepton physics. Therefore, we have
\begin{align}
    P(C_\ell^\obs \rvert \xi_\alpha, \omega_b) &= P(C_\ell^\obs \rvert \omega_b, \Neffxi^\code, {\Yp}^\code_{(\xi_\alpha,\omega_b)}) \nonumber \\
    &= \L_\mathrm{CMB}\left(\omega_b,\Neffxi^\code, {\Yp}^\code_{(\xi_\alpha,\omega_b)} \right) \, .
\end{align}
The CMB likelihood is expressed thanks to the vector $\vec{C} = (\omega_b, \Neff, \Yp)^T$ and the covariance matrix $\Sigma$ such that:\footnote{Note that the numerical uncertainty on $\Yp^\code$ is much smaller than the standard deviation of the CMB posterior distribution of $\Yp$ and is thus neglected here.}
\begin{multline}
\label{eq:likelihood_CMB}
    \L_\mathrm{CMB}(\omega_b,\Neff,\Yp) \\ = \frac{1}{(2 \pi)^{3/2} \sqrt{\abs{\mathrm{det} \, \Sigma}}} e^{- (\vec{C}-\vec{C}^\mathrm{obs})^T \cdot \Sigma^{-1} \cdot (\vec{C}-\vec{C}^\mathrm{obs})/2} \, .
\end{multline}
The central values and covariance matrices are given in Eqs.~\eqref{eq:means_Planck}--\eqref{eq:cov_allCMB}. They are obtained with the MCMC sampler \texttt{Cobaya}~\cite{Torrado:2020dgo}; see Fig.~\ref{fig:triangle_CMB}.

We finally express the total likelihood~\eqref{eq:likelihood_withom}, given by the product of individual likelihoods since measurements are independent: 

\begin{multline}
\label{eq:app_likelihood_tomarg}
\L(\xi_\alpha, \omega_b \rvert C_\ell^\obs,\Yp^\obs,\DH^\obs) \\
= \L_\mathrm{CMB}\left(\omega_b, \Neffxi^\code, {\Yp}^\code_{(\xi_\alpha,\omega_b)} \right) \\
\times \mathcal{N}\left(\Yp^\obs ; {\Yp}^\code_{(\xi_\alpha,\omega_b)}, \sigma_{\Yp}\right) \\ \times \mathcal{N}\left(\DH^\obs ; {\DH}^\code_{(\xi_\alpha,\omega_b)}, \sigma_{\DH}\right) \, .
\end{multline}

\section{Thermodynamic quantities}
\label{App:thermo}

Let us consider a given species with energy density $\rho$, pressure $P$, temperature $T$, number density $n$, and chemical potential $\mu$. Its volume entropy is given by
\begin{equation}
s  = \frac{\rho+P - n \mu}{T} \, .
\end{equation}
In the following, we give useful expressions for the various thermodynamic quantities of the relativistic species in the early Universe.

\paragraph*{Photons —} They have a Bose-Einstein distribution with zero chemical potential, hence
\begin{subequations}
\begin{align}
\rho_\gamma &= (z\Tcm )^4 \frac{\pi^2}{15}\, , \\ 
P_\gamma &= \rho_\gamma/3\, , \\ 
s_\gamma &= (z\Tcm)^3 \frac{4\pi^2}{45} \, .
\end{align}
\end{subequations}

\paragraph*{Electrons and positrons —} The charged leptons follow a Fermi-Dirac distribution with two helicity states. As long as they are ultrarelativistic ($T_\gamma \gg m_e$), we can neglect the very small asymmetry and we have
\begin{subequations}
\begin{align}
    \rho_{e^\pm} &= (z \Tcm)^4 \frac{7 \pi^2}{60} \, ,\\
    P_{e^\pm} &= \rho_{e^\pm}/3 \, , \\
    s_{e^\pm} &= (z \Tcm)^3 \frac{7 \pi^2}{45} \, .
\end{align}
\end{subequations}

\paragraph*{(Anti)neutrinos —} Their equilibrium distributions are Fermi-Dirac with only one helicity state, hence the total densities (for the sum of neutrinos and antineutrinos with $\bar \mu_\alpha = - \mu_\alpha$) are
\begin{subequations}
\begin{align}
\rho_\alpha + \bar{\rho}_\alpha &= (z_\alpha \Tcm)^4 \left(\frac{7 \pi^2}{120} +
  \frac{\xi_\alpha^2}{4} + \frac{\xi_\alpha^4}{8\pi^2}\right)\,, \\
  P_\alpha &= \rho_\alpha/3\,, \\ 
  s_\alpha + \bar{s}_\alpha &= (z_\alpha\Tcm)^3\left(\frac{7\pi^2}{90} + \frac{\xi_\alpha^2}{6}\right) \, .
\end{align}
\end{subequations}

\section{Transition from the \ATAOJH to the \ATAOH scheme}
\label{App:Ttrans}

As discussed in Sec.~\ref{subsec:Nu_transport}, we solve the QKEs using different schemes depending on the temperature. Notably, for low enough temperatures, the self-interaction mean-field Hamiltonian does not drive the dynamics anymore and can be discarded, which corresponds to the switch from the \ATAOJH to the \ATAOH scheme~\cite{Froustey2021}. The transition temperature between those schemes, $T_\trans$, is by default $\bar{T}_\trans = 2.2 \, \mathrm{MeV}$, but it is adapted in two ways. First, it is rescaled by a proxy for the photon temperature due to asymmetry equilibration, $z_\eqb$, in order to track the same range of $T_\gamma=z \Tcm$, since this is the relevant temperature for neutron-to-proton freeze-out. Then, it is further decreased for particular cases where oscillations are delayed (see Sec.~\ref{subsec:EBO}). In this Appendix, we show the necessity for such an adaptive method, and quantify the errors incurred if the transition is made too early.

\begin{figure*}[!ht]
    \centering
    \includegraphics{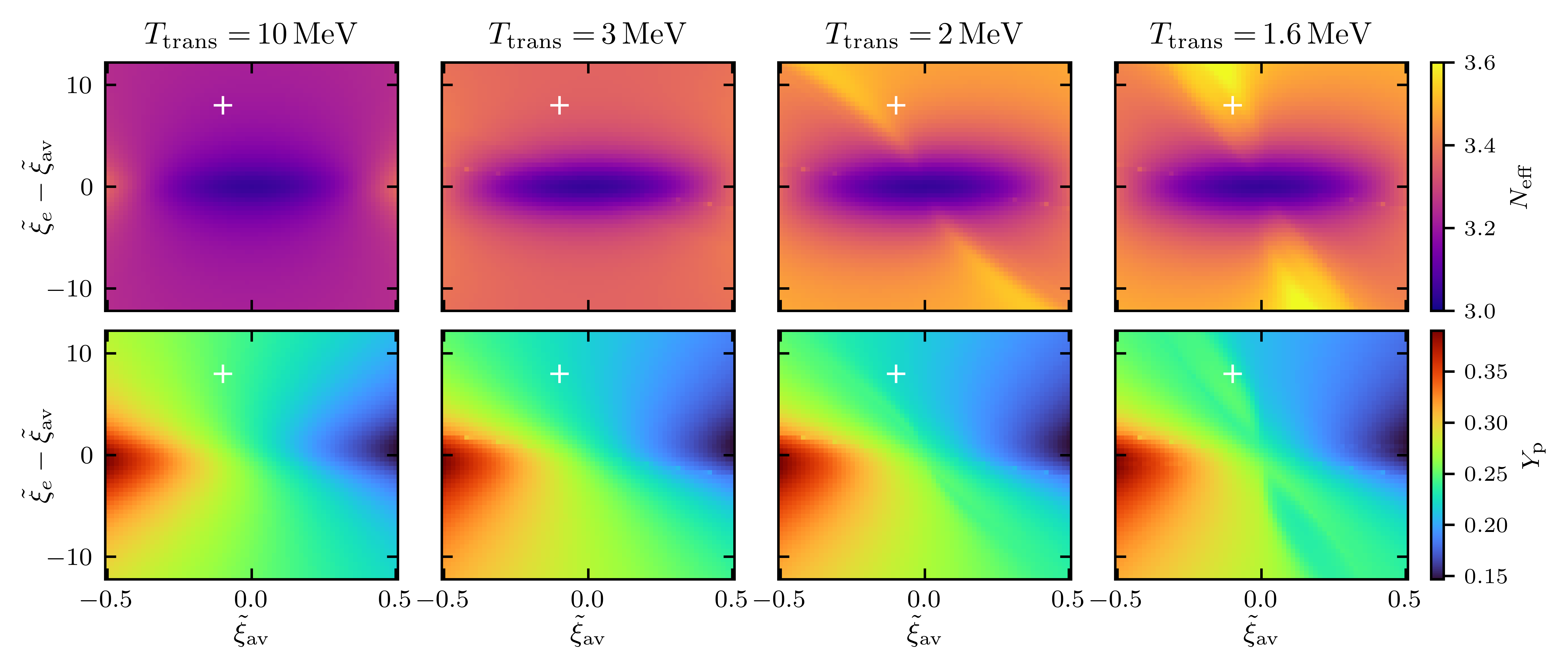}
    \caption{\label{fig:output_Ttrans} Final $\Neff$ (top panel) and helium-4 abundance (bottom panel) for initial asymmetries in the plane $(\xit_\av,\delta \xit_e)$, for different transition temperatures $T_\trans$. The precise results, obtained with the adaptive method, are shown on Fig.~\ref{fig:output_xie}. The point with coordinates $(-0.1,8.0)$ is shown with a white cross, and it is studied specifically in Figs.~\ref{fig:NeffYp_Ttrans} and \ref{fig:evolution_Ttrans}.}
\end{figure*}

To illustrate the impact of the choice of $T_\trans$, we focus on initial asymmetries in the plane $(\xit_\av,\delta \xit_e)$ with $\Deltat = 0$, a situation studied in Sec.~\ref{subsec:xie}. The results of \nevo, using the adaptive setting of $T_\trans$, are shown on Fig.~\ref{fig:output_xie}, and we show how a different $T_\trans$ changes those outputs on Fig.~\ref{fig:output_Ttrans}. Generally speaking, a too large $T_\trans$ can lead to errors of order $10 \, \% - 20 \, \%$. However, this is highly dependent on the initial asymmetries: for small values of $(\xit_\av, \delta \xit_e)$, we see on the top panel that the smaller value of $\Neff$ is well described regardless of $T_\trans$. The “polar” regions on Fig.~\ref{fig:output_Ttrans} show particular sensitivity to $T_\trans$, as one needs to go below $2 \, \mathrm{MeV}$ to start discerning the patterns found on Fig.~\ref{fig:output_xie}.

\begin{figure}[!ht]
    \centering
    \includegraphics{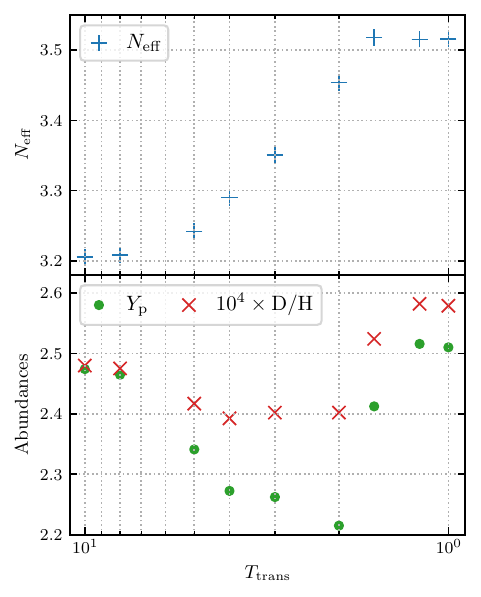}
    \caption{\label{fig:NeffYp_Ttrans} Final value of $\Neff$ and helium-4 and deuterium abundances from the initial asymmetries $(\xit_\av = - 0.1, \, \delta \xit_e = 8.0, \, \Deltat = 0.)$, for different values of the transition temperature $T_\trans$.}
\end{figure}

The trends in variations of $\Neff$ and $\Yp$ can be understood by looking at a specific example. We focus in the following on the initial asymmetries identified by a white cross on Fig.~\ref{fig:output_Ttrans}: ($\xit_e = 7.9$, $\xit_\mu = \xit_\tau = -4.1$). The variation of the final values of $\Neff$, $\Yp$ and $\DH$ for this initial configuration as a function of $T_\trans$ are shown on Fig.~\ref{fig:NeffYp_Ttrans}. We observe a convergence of the values when $T_\trans$ approaches $1 \, \mathrm{MeV}$. If $T_\trans$ is too large, $\Neff$ is underestimated while the abundances show a more complicated pattern: as $T_\trans$ diminishes, the abundances decrease before increasing back until they converge to the actual values.

\begin{figure*}[!ht]
    \centering
    \includegraphics{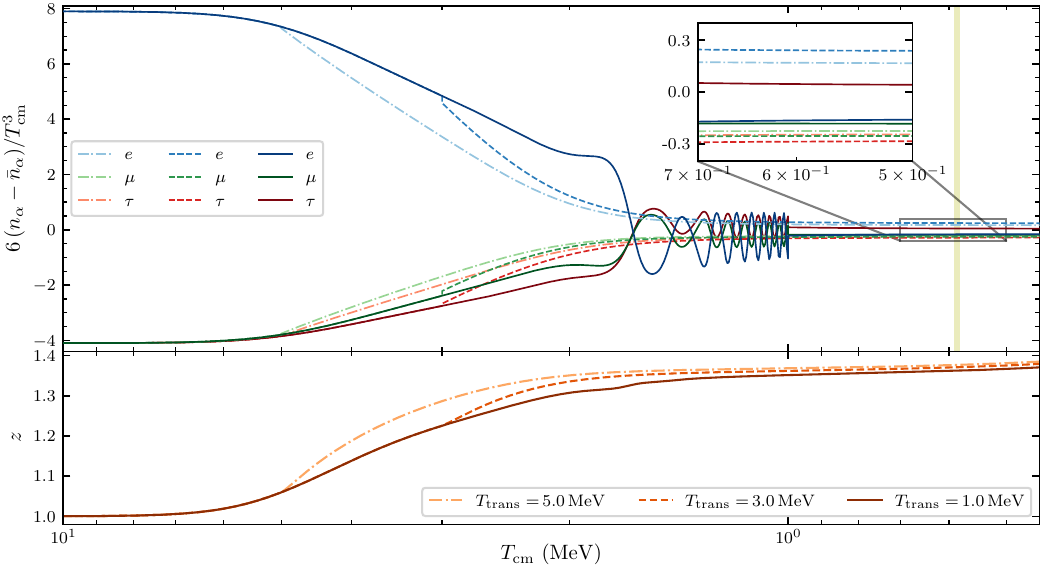}
    \caption{\label{fig:evolution_Ttrans} Evolution of neutrino asymmetries (top panel) and the photon comoving temperature (bottom panel), for the initial configuration $(\xit_\av = - 0.1, \delta \xit_e = 8.0, \Deltat = 0.)$. Three different transition temperatures are shown: $T_\trans = 5 \, \mathrm{MeV}$ (dash-dotted lines, light shades), $T_\trans = 3 \, \mathrm{MeV}$ (dashed lines, medium shades) and $T_\trans = 1 \, \mathrm{MeV}$ (solid lines, dark shades).}
\end{figure*}

We interpret the behavior of $\Neff$ in the same way as in the discussion of Sec.~\ref{subsec:EBO}. The mixing of asymmetries is associated to a transfer of energy from the degenerate sea of (anti)neutrinos to the ensemble \{neutrinos+antineutrinos+electromagnetic plasma\}. Imposing a higher value of $T_\trans$ artificially makes this energy redistribution happen earlier. This can be seen on Fig.~\ref{fig:evolution_Ttrans}, where we show the time evolution of asymmetries and $z$: when $T_\trans$ is too high ($5 \, \mathrm{MeV}$ and $3 \, \mathrm{MeV}$ shown on this figure), $z$ gets an unphysical “bump” at $T_\trans$. Now, the higher $T_\trans$, the more coupled neutrinos are with the electromagnetic plasma, and the less neutrinos are comparatively more reheated than photons by the mixing of asymmetries. In other words, the higher $T_\trans$, the lower $\Neff$ as seen on Fig.~\ref{fig:NeffYp_Ttrans}. For $T_\trans \leq 1.6 \, \mathrm{MeV}$, $\Neff$ asymptotes to its actual value: as we can see on Fig.~\ref{fig:evolution_Ttrans}, this is because the reduction of the asymmetries has been properly described by the \ATAOJH scheme, and we are only left with low-amplitude very fast collective oscillations that we average by switching to the \ATAOH scheme. Regarding the abundances, we see on Fig.~\ref{fig:evolution_Ttrans} (see in particular the inset plot) that the final value of $\eta_e$ first increases with decreasing $T_\trans$ (from $5 \, \mathrm{MeV}$ to $3 \, \mathrm{MeV}$), before decreasing to its actual value. This is associated to a reduction then a rise in the neutron-to-proton ratio, hence a reduction then a rise of the abundances, consistently with Fig.~\ref{fig:NeffYp_Ttrans}. We see on this example that $T_\trans$ must be below $2 \, \mathrm{MeV}$ to capture the synchronous oscillations and provide the accurate final value of $\eta_e$.

In general, various configurations require different values of $T_\trans$, and we extensively checked that our adaptive scheme meets this requirement. There are still particularly challenging points corresponding to equal-but-opposite configurations, discussed in Sec.~\ref{subsec:EBO}.

\section{Role of deuterium}
\label{App:deuterium}

Nuclear rates represent a major source of uncertainty for the prediction of primordial abundances from BBN. In particular, depending on the determination of the rates of $\mathrm{D(d,n)^{3}He}$ and $\mathrm{D(d,p)^{3}He}$ reactions, the deuterium abundance calculated for vanishing asymmetries is found to be either fully consistent~\cite{Parthenope_revolutions,Yeh2020} or in mild tension~\cite{Pitrou2020} with the experimental measurement~\eqref{eq:DH_mes}. Detailed studies of the differences between the \parthenope and \primat codes have confirmed that the disagreement, which stems from different methods and data selections to obtain the nuclear rates at the BBN scale, can only be resolved with future, more accurate, nuclear rate measurements~\cite{Pisanti2020,Pitrou2021}. In the meantime, our use of \primat comparatively favors configurations with a final $\xi_e \lesssim 0$, such that the deuterium abundance is increased toward~\eqref{eq:DH_mes}. 

However, this effect is subdominant and does not significantly alter our conclusions: to illustrate the role played by the deuterium abundance, we run the same analysis as in Sec.~\ref{subsec:xie} but removing the likelihood from deuterium measurements. In other words, the likelihood~\eqref{eq:likelihood_tomarg} becomes

\begin{multline}
\label{eq:likelihood_noD}
\L(\xi_\alpha, \omega_b \rvert C_\ell^\obs,\Yp^\obs)
= \L_\mathrm{CMB} \left(\omega_b, \Neffxi^\code, {\Yp}^\code_{(\xi_\alpha,\omega_b)} \right) \\
\times \mathcal{N}\left(\Yp^\obs ; {\Yp}^\code_{(\xi_\alpha,\omega_b)}, \sigma_{\Yp}\right) \, .
\end{multline}

\begin{figure}[!ht]
    \centering
    \includegraphics{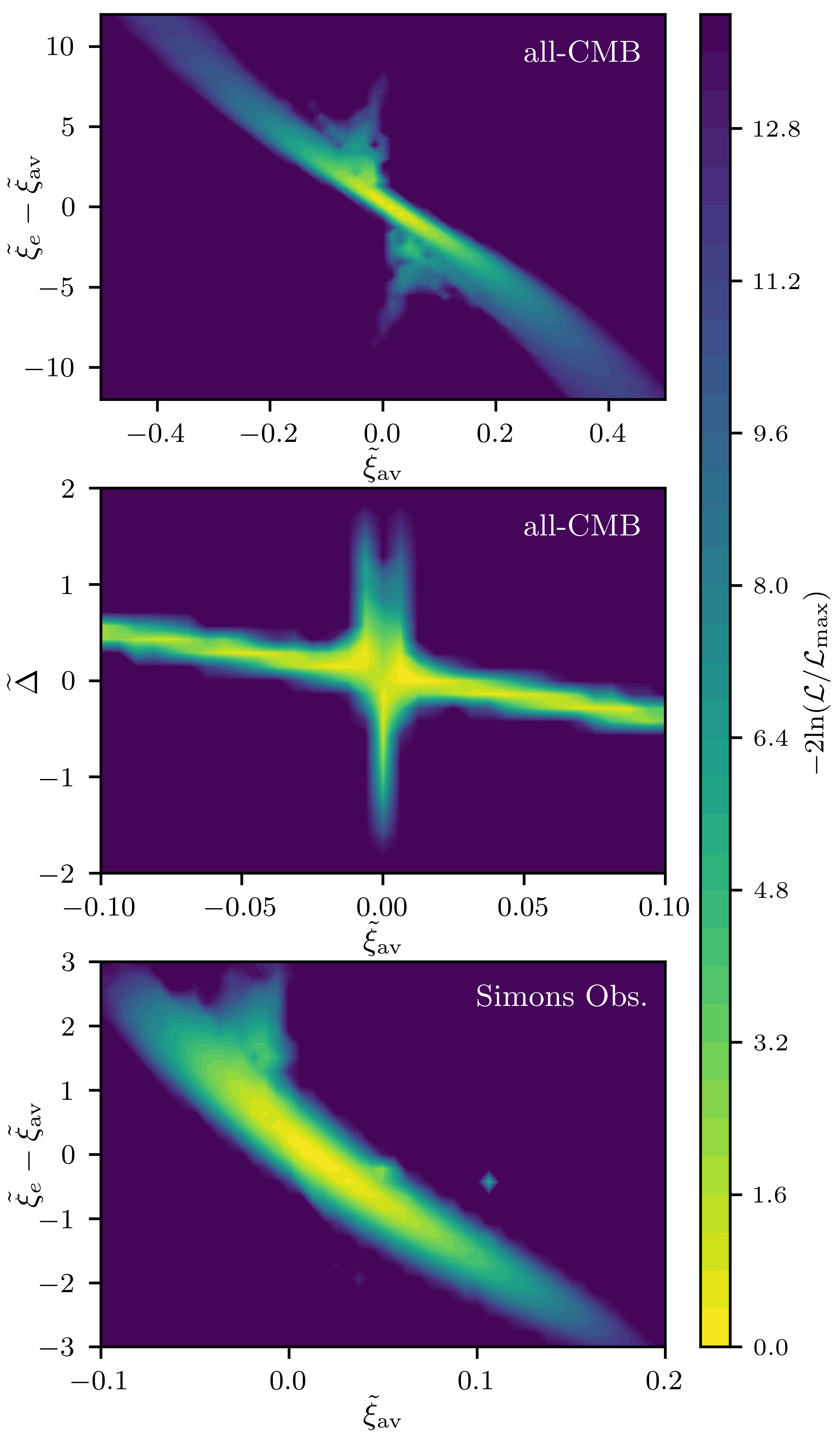}
    \caption{\label{fig:likelihood_noD} Likelihood of primordial asymmetries without taking into account the measurement of $\DH$. The helium-4 abundance value is taken from Aver et al., Eq.~\eqref{eq:Yp_mes}. \emph{Top:} results in the plane $(\xit_\av, \delta \xit_e)$, to be compared with Fig.~\ref{fig:likelihood_all_xie} (bottom left panel). \emph{Middle:} results in the plane $(\xit_\av, \Deltat)$, to be compared with Fig.~\ref{fig:likelihood_delta}. \emph{Bottom:} forecast of Simons Observatory constraints, using $\Yp\rvert_\mathrm{Aver}$ as the central value in $\mathcal{L}_\mathrm{CMB}$, to be compared with Fig.~\ref{fig:likelihood_forecast_xie_Ypmean} (top panel).}
\end{figure}

The associated results are shown on Fig.~\ref{fig:likelihood_noD}.

On the top panel, we see that the space of allowed asymmetries is more constrained (compare with Fig.~\ref{fig:likelihood_all_xie}, bottom left panel) when the deuterium constraints are not taken into account. This counterintuitive property is due to the very non-Gaussian features of the joint likelihood. In particular, because of the tension between the $\DH$ value predicted by \primat at zero asymmetries and the measurement~\eqref{eq:DH_mes}, regions away from the point $(0,0)$ are favored at a comparable level with the central region. In the plane $(\xit_\av,\Deltat)$ (Fig.~\ref{fig:likelihood_noD}, central panel), the high likelihood regions are still unbound. Finally, the bottom panel is almost indistinguishable from the top panel of Fig.~\ref{fig:likelihood_forecast_xie_Ypmean}: it is a consequence of the higher constraining power of this future CMB likelihood, which makes the mild deuterium tension only marginally relevant.

Our main conclusion is thus unaffected: the space of allowed asymmetries is generically unbound with current CMB data, but future experiments should efficiently constrain the possible values of $\xi_\alpha$. This nevertheless highlights that future spectroscopic measurements of the deuterium abundance, which will reduce the uncertainty levels, might play an important role in future constraints — which shows how timely better nuclear rate measurements are, as they could become, by large, the main source of uncertainty in the BBN likelihood.

\bibliographystyle{bibi}
\bibliography{BiblioAsymmetries}

\end{document}